\documentclass[conference]{IEEEtran}

\usepackage{graphicx}                % allow us to embed graphics files

\graphicspath{{figures/}{pictures/}{images/}{./}} % where to search for the images

\usepackage{microtype}                 % use micro-typography (slightly more compact, better to read)
\PassOptionsToPackage{warn}{textcomp}  % to address font issues with \textrightarrow
\usepackage{textcomp}                  % use better special symbols
\usepackage{mathptmx}                  % use matching math font
\usepackage{times}                     % we use Times as the main font
\usepackage{cite}                      % needed to automatically sort the references
\usepackage{tabu}                      % only used for the table example
\usepackage{booktabs}                  % only used for the table example
\usepackage{tabularx}

\usepackage{booktabs}
\newcommand{\ra}[1]{\renewcommand{\arraystretch}{#1}}

\usepackage{multirow}

\usepackage{subcaption}
\usepackage{nameref}

\usepackage[table]{xcolor}

\usepackage{amstext}

\usepackage{url}

\usepackage{enumitem}

\renewcommand{\baselinestretch}{0.99}

\begin{document}

%
% paper title
% Titles are generally capitalized except for words such as a, an, and, as,
% at, but, by, for, in, nor, of, on, or, the, to and up, which are usually
% not capitalized unless they are the first or last word of the title.
% Linebreaks \\ can be used within to get better formatting as desired.
% Do not put math or special symbols in the title.
\title{Code Park: A New 3D Code Visualization Tool}

% author names and affiliations
% use a multiple column layout for up to three different
% affiliations
\author{\IEEEauthorblockN{Pooya Khaloo\IEEEauthorrefmark{1}, Mehran Maghoumi\IEEEauthorrefmark{2}, Eugene Taranta II\IEEEauthorrefmark{3}, David Bettner\IEEEauthorrefmark{4}, Joseph Laviola Jr.\IEEEauthorrefmark{5}}
\IEEEauthorblockA{University Of Central Florida\\
Email: \IEEEauthorrefmark{1}pooya@cs.ucf.edu, \IEEEauthorrefmark{2}mehran@cs.ucf.edu, \IEEEauthorrefmark{3}etaranta@gmail.com, \IEEEauthorrefmark{4}dbettner@gmail.com, \IEEEauthorrefmark{5}jjl@cs.ucf.edu	}
}

%% The ``\maketitle'' command must be the first command after the
%% ``\begin{document}'' command. It prepares and prints the title block.

% make the title area
\maketitle

% As a general rule, do not put math, special symbols or citations
% in the abstract
\begin{abstract}
We introduce Code Park, a novel tool for visualizing codebases in a 3D game-like environment. Code Park aims to improve a programmer's understanding of an existing codebase in a manner that is both engaging and intuitive, appealing to novice users such as students. It achieves these goals by laying out the codebase in a 3D park-like environment. Each class in the codebase is represented as a 3D room-like structure. Constituent parts of the class (variable, member functions, \textit{etc.}) are laid out on the walls, resembling a syntax-aware ``wallpaper". The users can interact with the codebase using an overview, and a first-person viewer mode. We conducted two user studies to evaluate Code Park's usability and suitability for organizing an existing project. Our results indicate that Code Park is easy to get familiar with and significantly helps in code understanding compared to a traditional IDE. Further, the users unanimously believed that Code Park was a fun tool to work with.
\end{abstract}

% For peer review papers, you can put extra information on the cover
% page as needed:
% \ifCLASSOPTIONpeerreview
% \begin{center} \bfseries EDICS Category: 3-BBND \end{center}
% \fi
%
% For peerreview papers, this IEEEtran command inserts a page break and
% creates the second title. It will be ignored for other modes.
\IEEEpeerreviewmaketitle

%% the only exception to this rule is the \firstsection command
\section{Introduction}

%% \section{Introduction} %for journal use above \firstsection{..} instead

Code visualization techniques assist developers in gaining insights into a codebase that may otherwise be difficult or impossible to acquire via the use of a traditional text editor. Over the years, various techniques have been proposed to address different issues. For instance, SeeSoft \cite{eick1992seesoft} maps each line of code to an auxiliary UI bar such that the color of the bar represents a property of the code, \textit{e.g.} red rows could represent recently changed lines. Code Bubbles \cite{bragdon2010codebubble} organizes source into interrelated editable fragments illustrated as bubbles within the user interface, and CodeCity \cite{wettel2008codecity2} uses a 3D city-like structure to visualize the relative complexity of modules within a codebase. However, among all of the techniques that have been explored, little attention has been given to how source code can be visualized in order to help developers become familiar with and learn a \emph{new} codebase.

%Understanding an existing codebase is often a huge undertaking: a programmer has to comb through hundreds of files that are scattered everywhere to understand how somebody else’s code works. Code understanding and remembering code structure becomes even more important when a programmer wants to debug an existing piece of software. Newcomers to the field of software development quickly discover that they spend vast majority of their time trying to understand and learn an existing codebase. The user’s understanding usually comes from studying and reading the code. However, this form of understanding can quickly become tedious and boring, because human information-processing system has limited capacity \cite{baddeley1997human}. This possibly leads to discouraging beginners from trying to learn by reading code. This issue can be mitigated with code visualizers that display constituent part of a codebase in a more clear and coherent manner, such as  \cite{bragdon2010codebubble, yang2003solar, wettel2007codecity}. Each of these tools attempt to show as much information as possible from a codebase in one frame.  They use different methods to help user understand and remember the structure of a codebase more efficiently. However, they are mostly unable to show the details of each constituent part and do not focus on encouraging user engagement.

\begin{figure}
\centering
\begin{subfigure}{0.95\columnwidth}
  \includegraphics[width=\columnwidth]{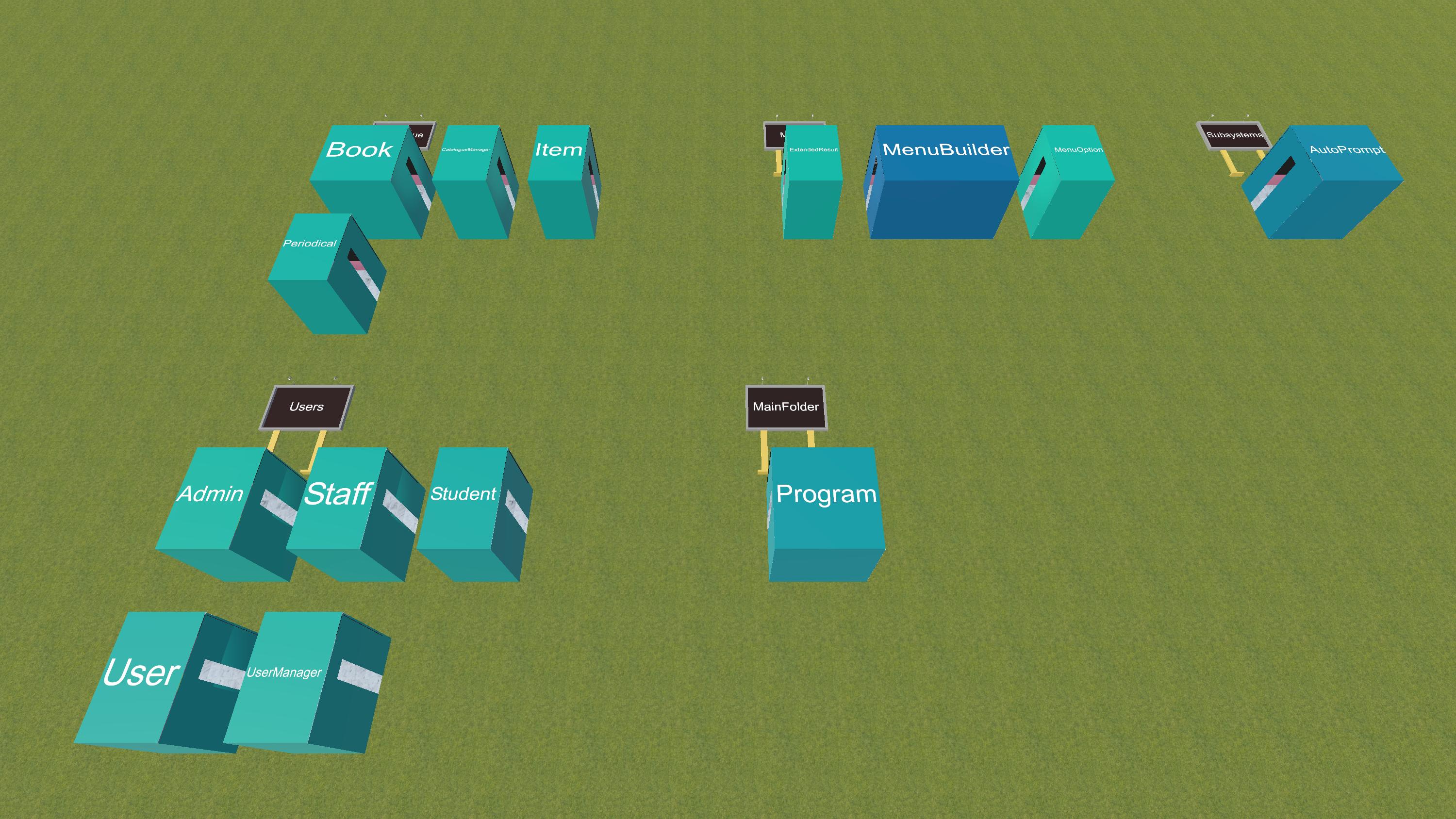}
\caption{Top-down view of the entire codebase. Each cube is a code room.} \label{fig:InitialView}
\end{subfigure}
\vspace*{\fill} % separation between the subfigures
\begin{subfigure}{0.95\columnwidth}
  \includegraphics[width=\columnwidth]{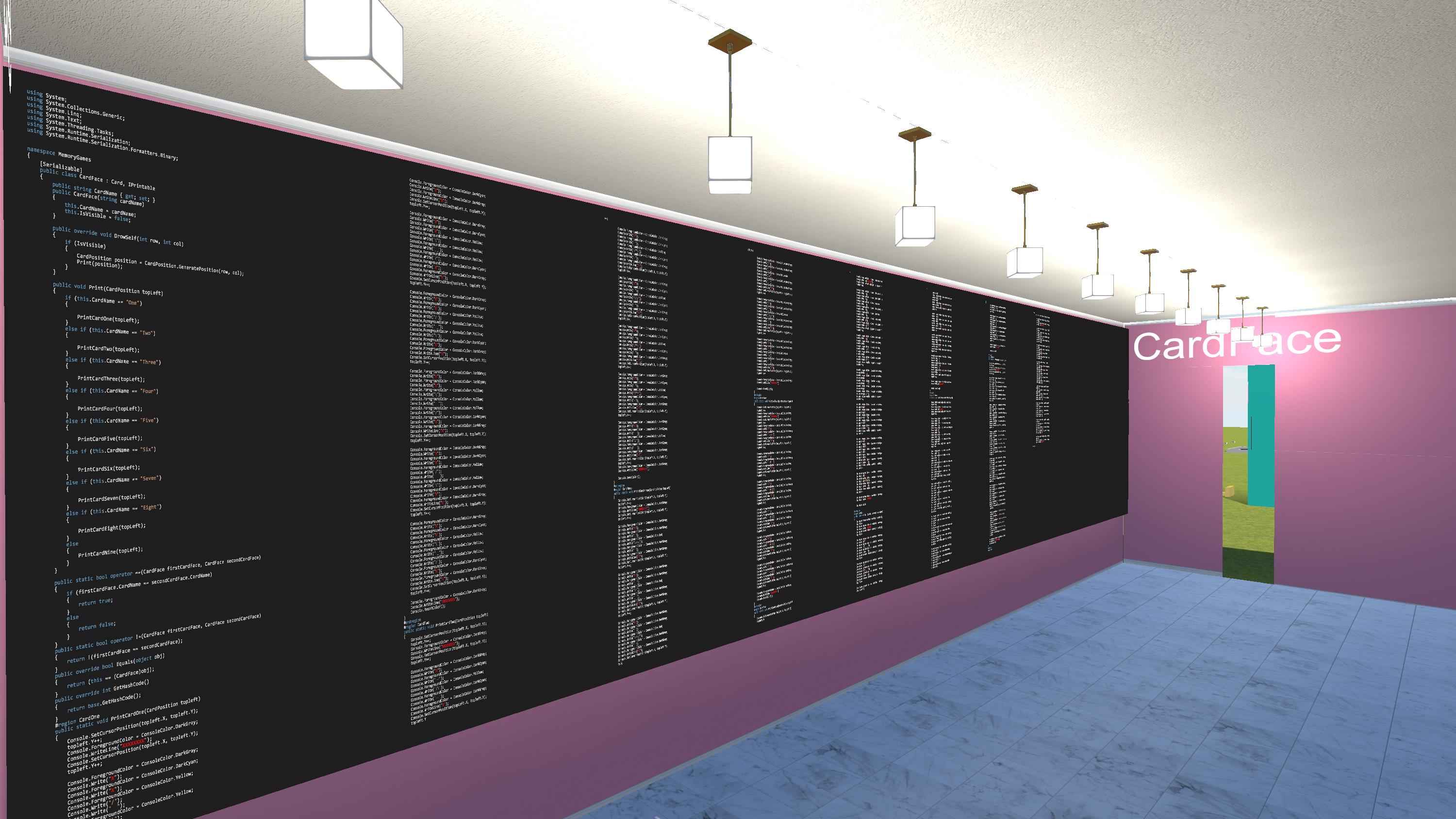}
\caption{Exploring the inside of a code room in first-person vieweing mode.} \label{fig:BigCode}
\end{subfigure}
\caption{Code Park in action. Each class is shown as a room. The user can study the code in a first-person view mode.} \label{fig:Fig1}
%\vspace{-5mm}
\end{figure}

The task of learning a new codebase is fraught with many issues, three of which include memorization, cognitive load, and engagement \cite{baddeley1997human,SWELLER1994CognitiveLoad,Carini2006Engagement}. Imagine a new hire set to study a proprietary project that is a complex hierarchical conglomeration of thousands of source files. Indeed memorizing the structure of such a system is a difficult chore, not to mention mentally exhausting as well as potentially tedious and likely boring. To address these issues, we introduce Code Park (CP), a new 3D code visualization tool that aims to improve a programmer's understanding of an existing codebase in a manner that is both engaging and fun. CP organizes source code into a 3D scene in order to take advantage of human's spatial memory capabilities \cite{burgess2002human} and help one better understand and remember the architecture. CP also supports two points of view that are an exo-centric (bird's eye) and ego-centric (first-person) view, which allows one to examine the codebase at different granularities (see Figure \ref{fig:Fig1}).

%To mitigate these issues, we present Code Park (CP), a new 3D code visualization tool that aims to improve the programmer’s understanding of an existing codebase in a manner that is both engaging and fun. By laying out the constituent parts of a codebase in the 3D space, CP encourages engagement with the code. As humans, we tend to easily remember the position of the objects that we work with using our 3D spatial memory  \cite{burgess2002human}. CP taps into this affordance to help with understanding and remembering the code. Further, CP provides two levels of code interaction: exo-centric and ego- centric views (see Figure \ref{fig:Fig1}), which make examining the code at different granularities possible. At the time of this writing, CP only supports C\# codebases. Nevertheless, it can be trivially extended to any programming language that supports object-oriented programming.

Our design goals with CP are threefold. We aimed to create a code visualization tool that helps users become familiar with and memorizing an existing codebase. This tool must be easy to work with and must show information in a form which reduces the user's cognitive load. Finally, the tool must be engaging and enjoyable to work with. Specifically the following are the contributions of our exploratory work:
%Our motivation was to explore a novel means of interacting with code to gauge user interest and examine code understanding in 3D game-like environment. We aimed to create an easy to use tool that can show information in a way that reduces the user's cognitive load. Specifically the following are the contributions of our exploratory work:

\begin{enumerate}
	\item Creating an engaging and intuitive tool for improved code understanding.
	\item Examining the usability of such a tool and gauging user interest.
	\item Displaying the source code itself in a 3D environment (rather than a metaphorical representation) and facilitate ``intimate" interaction with the code via an ego-centric mode.
\end{enumerate}

To evaluate our design goals and the usability of the system, we performed a user study. Our results indicate that our participants found CP easy to use and helpful in code understanding. Additionally, they unanimously believed that CP was enjoyable. We ran a follow up user study to determine how users would organize the classes of an existing project in the 3D environment in a way that the arrangement helps them remember the code structure better.

%This paper is organized as follows. We first examine some of the related work available in the literature. We then proceed with the discussion of our user interface and the reason for our design choices. Next, we present our user study design and usability evaluation and provide an in-depth statistical analysis of our results. After that, we present the limitations that are associated with our current system and discuss possible future work to address those shortcomings and further improve our work. 

%\begin{figure*}
%\centering
%\begin{subfigure}{0.33\textwidth}
%  \includegraphics[width=\columnwidth]{figures/OldCodeHouse1.jpg}
%\caption{The house's exterior.} \label{fig:OldCodeHouseView}
%\end{subfigure}
%%\hspace*{\fill} % separation between the subfigures
%\begin{subfigure}{0.33\textwidth}
%  \includegraphics[width=\columnwidth]{figures/OldCodeHouse2.jpg}
%\caption{The living room with class files on the walls.} \label{fig:OldCodeHouseLivingRoom}
%\end{subfigure}
%%\hspace*{\fill} % separation between the subfigures
%\begin{subfigure}{0.33\textwidth}
%  \includegraphics[width=\columnwidth]{figures/OldCodeHouse3.jpg}
%\caption{A closer view of the code on the wall.} \label{fig:OldCodeHouseBedroom}
%\end{subfigure}
%\caption{The first prototype of CP.} \label{fig:OldCodeHouse}
%\end{figure*}

%----------------------------------------------------------------------------------------
%	Related Work
%----------------------------------------------------------------------------------------
\section{Related Work}

%In the past decades many researchers have shown the difficulties associated with text-based 2D environments \cite{ko2006exploratory, murphy2006java, robillard2004effective, sherwood2008path}. These studies have shown that developers spend most of their time on navigation to understand the code.
 There is a body of work available in the literature on software visualization \cite{balzer2007lvl,bonyuet2004software,vissoft2,vissoft1,greevy2005feature,vissoft3,vissoft4}. We direct the reader to comprehensive surveys of different methods available in the work of Teyseyre and Campo \cite{survey2009} and also Caserta and Olivier \cite{caserta2011survey}. For example SeeSoft \cite{eick1992seesoft}, one of the earliest visualization metaphors, allows one to analyze up to 50,000 lines of code by mapping each line of code into a thin row. Marcus \textit{et al.} \cite{marcus20033d} added a new dimension to SeeSoft to support an abstraction mechanism to achieve better representation of higher dimensional data.

Recently, there have been more work focusing on 2D visualization. Code Bubbles \cite{bragdon2010codebubble} suggested a collection of editable fragments that represent functions in a class. Code Gestalt \cite{Kurtz2011Gestalt} used tag overlay and thematic releations. Lanza and Ducasse \cite{lanza2001categorization} proposed categorizing classes and their internal objects into blocks called Blue Prints. Gutwenger \textit{et al.} \cite{gutwenger2003new} proposed an approach for improved aesthetic properties of UML diagrams when visualizing hierarchical and non-hierarchical relations. Balzer \textit{et al.} \cite{balzer2005voronoi} introduced hierarchy-based visualization for software metrics using Voroni Treemaps. Additionally, Holten \cite{holten2006hierarchical} used both hierarchical and non-hierarchical data to visualize adjacency relation in software. The common observation among these efforts is that they are all based on 2D environments and were mostly suitable for expert users. 

%\subsection{Why prefer 3D over 2D?}
\vspace{2mm}
\noindent
\textbf{Benefit of 3D over 2D.~}
Remembering code structure will result in faster development so it is an essential part of being a programmer. Specifically, 3D environments tap into the spatial memory of the user and help with memorizing the position of objects \cite{burgess2002human}. These objects could be classes or methods. There are also studies which provide evidence that spatial aptitude is a strong predictor of performance with computer-based user interfaces. For instance, Cockburn and McKenzie \cite{Cockburn2002Spatial} have shown that 3D interfaces that leverage the human's spatial memory result in better performance even though some of their subjects believed that 3D interfaces are less efficient. Robertson \textit{et al.} \cite{Robertson1998DataMountain} have also shown that spatial memory does in fact play a role in 3D virtual environments. A number of researchers have attempted to solve the problem of understanding code structure. Graham \textit{et al.} \cite{yang2003solar} suggested a solar system metaphor, in which each planet represented a Java class and the orbits showed various inheritance levels. Balzer \textit{et al.} \cite{balzer2004land} presented the static structure and the relation of object-oriented programs using 3D blocks in a 2D landscape model.

Among these, the city metaphor is one of the most popular ideas. CodeCity \cite{ wettel2008codecity2} is an example of a city metaphor. CodeCity focus on the visualization of large and complex codebases using city structures where the number of methods in each class represented the width of the buildings and the number of attributes represented their height. Panas \textit{et al.} \cite{panas2007vizz3d} used Vizz3D\footnote{\url{http://vizz3d.sourceforge.net/}} to visualize communication between stakeholders and developers in the process of software implementation. Alam and Dugerdil \cite{alam2007evospaces} introduced the Evospaces visualization tool in which they represent files and classes with buildings similar to CodeCity. Additionally, they showed the relations between the classes using solid pipes. There are other works that leverage such visualization metaphors such as \cite{knight2000virtual} and  \cite{langelier2005visualization}. The common theme with these tools is that most of them only showed the name of the classes in their environment which is not instrumental for learning purposes and again, they are mainly targeting experienced developers.

%\subsection{What about being engaging?}
\vspace{2mm}
\noindent
\textbf{User engagement.~}
The fun and engaging aspect of programming tools, especially for teaching purposes, has gained attention in recent years. Alice~\cite{Cooper2000Alice} is a 3D interactive animation environment tool that aims to encourage student engagement by creating a game-like environment. Resnik \textit{et al.} \cite{Resnick2009Scratch} introduced a novel development environment that appeals to people who believe their skills are not on par with experienced developers. They designed their tool in a way that was fun and likable. They also made it suitable for youth learners. There are also many studies that are focused on creating fun and engaging tools and methods that help newcomers enjoy learning programming \cite{Parsons2006Puzzle, kahn1996toontalk, Perrin2012algopath, Kelleher2007Storytelling}. 

What sets CP apart from current tools in the literature is that it is designed to be suitable for both beginner and experienced developers alike. Saito \textit{et al.} \cite{education} examined the learning effects between a visual and a text-based environment on teaching programming to beginners. Their results deemed the visual environment as a more suitable option for teaching beginners. Indeed, at the lines of code level, CP differs little from a traditional IDE, other than code can be read from within a 3D environment. Where CP diverges from a traditional IDE is in how one interacts with a project: file hierarchies vs buildings in 3-space. The exo-centric view in CP helps the user glean a holistic understanding of the codebase without involving them in unnecessary details. Conversely, the ego-centric view enables a fine-grained understanding of the codebase. The addition of the ego-centric view is the key distinction between the current work and CodeCity \cite{wettel2008codecity2}.

\begin{figure*}[t]
\centering
\begin{subfigure}{0.45\textwidth}
  \includegraphics[width=\columnwidth]{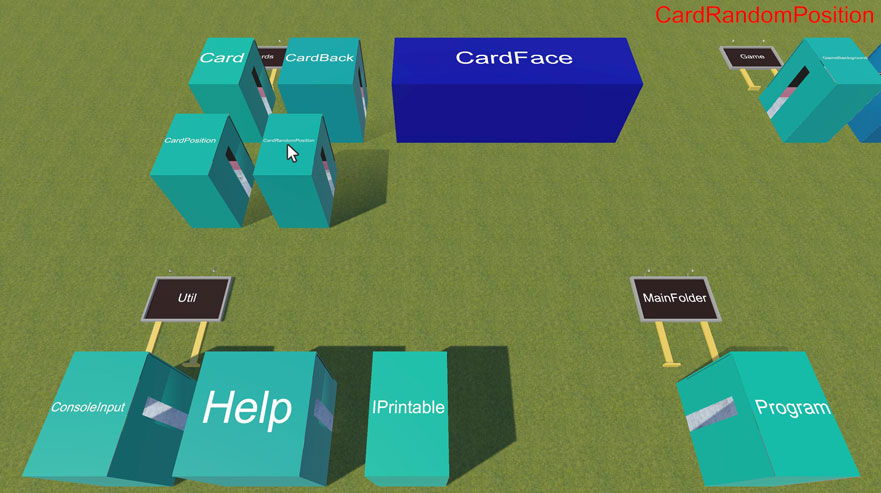}
\caption{Bird's view with tooltip on top-right.} \label{fig:ClassNameToolTip}
\end{subfigure}
\vspace*{\fill} % separation between the subfigures
\begin{subfigure}{0.45\textwidth}
  \includegraphics[width=\columnwidth]{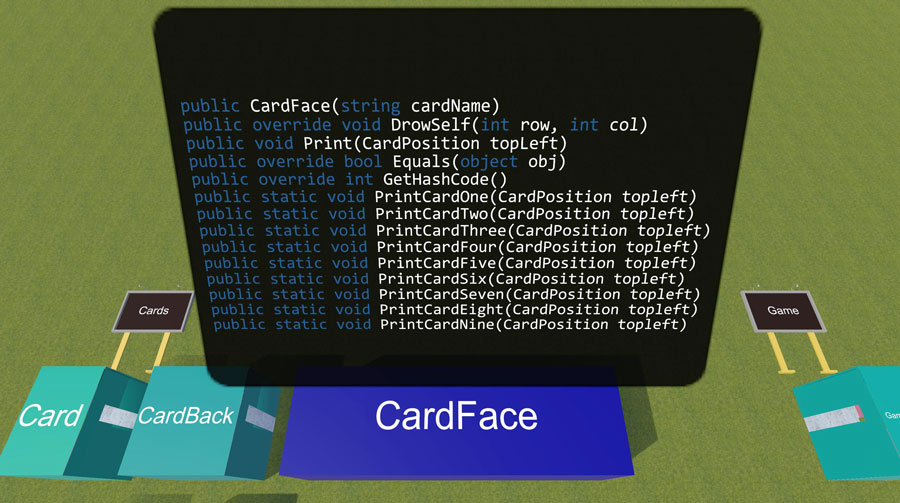}
\caption{Class method overview tooltip.} \label{fig:MethodNameToolTip}
\end{subfigure}
\begin{subfigure}{0.45\textwidth}
  \includegraphics[width=\columnwidth]{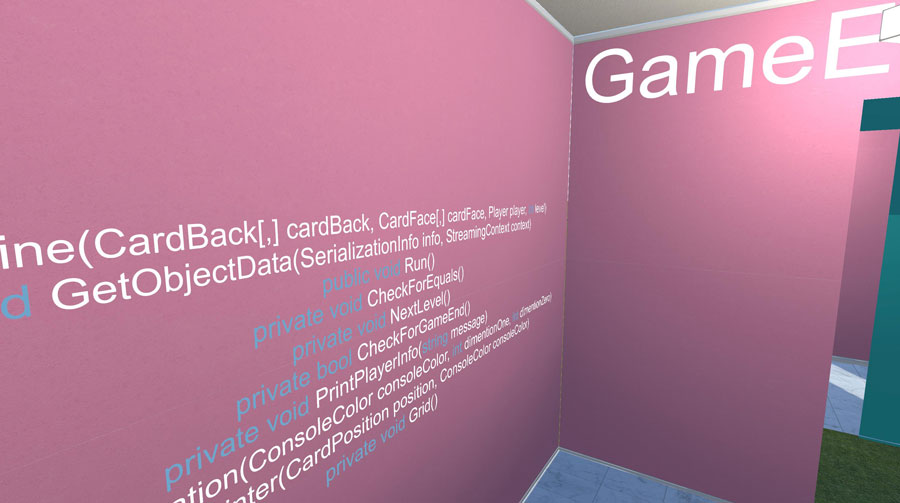}
\caption{Class method overview wallpaper.} \label{fig:WallMethod}
\end{subfigure}
\vspace*{\fill} % separation between the subfigures
\begin{subfigure}{0.45\textwidth}
  \includegraphics[width=\columnwidth]{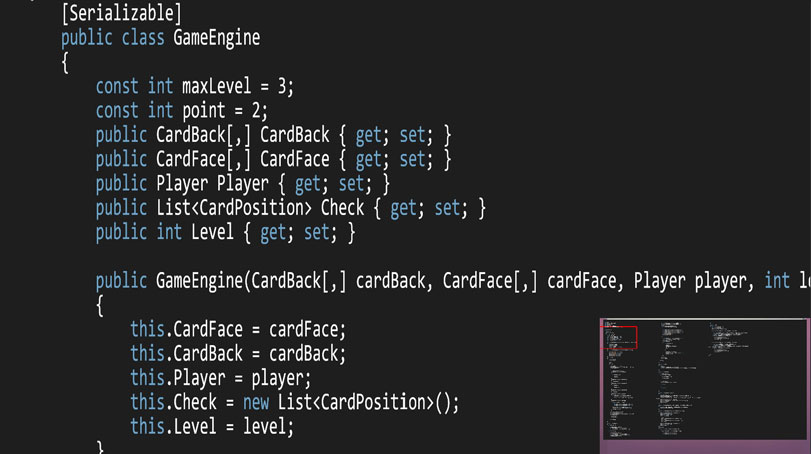}
\caption{Code reading view.} \label{fig:CodeBorder}
\end{subfigure}
\caption{Various CP view modes and features. In (a), the user has hovered the mouse over a class with a long name.} \label{fig:Fig2}
\end{figure*}

%----------------------------------------------------------------------------------------
%	IMPLEMENTATION
%----------------------------------------------------------------------------------------
\section{User Interface Design}
\label{sec:ui}

When designing CP, we had three main goals in mind. We wanted CP to greatly help with learning codebases, be easy to learn and fun to use. With these goals in mind, we employed an iterative design approach. We tested CP often and refined it by incorporating peer feedback from experts in this area.

In CP each room represents one class in the codebase. This approach is inspired by CodeCity \cite{wettel2008codecity2}.  The rooms are placed on a ground filled with a grassy texture, which resembles a park. The rooms are grouped together based on the code's project file structure: files inside the same directory result in adjacent rooms and each group of rooms are labeled by the directory name (see Figure \ref{fig:ClassNameToolTip}). This was done in order to avoid confusion for large codebases that had many files. The size and the color of each room is proportional to the size of the class they contain (larger classes appear as larger rooms that have a darker color on their exteriors). Each room has syntax-aware wallpapers that have a color theme matching Microsoft Visual Studio (VS)'s dark theme (dark background and gray text with colored language keywords, comments and strings literals -- see Figure \ref{fig:CodeBorder}). The users can explore the environment in the first-person mode because it is one of the well known navigation methods in the 3D environment and also resembles video games which helps making CP more engaging. The users can click on the wallpaper with the cross-hair to transition into the code viewing mode. This places the camera orthogonal to the wallpaper, allowing the users to read the code and scroll it (using the mouse wheel) similar to a text editor.

One of our design goals was to allow CP to show as much information as possible while maintaining the ease of use and managing the user's cognitive load. Consequently, we employed the \textit{bird's view} navigation mode (see Figure \ref{fig:ClassNameToolTip}). In bird view mode, the users have a top-down view of all the rooms placed on the ground. This way, the users can first get some insight about the structure of the codebase and the number of classes and then study the codebase in more detail via the first-person mode. In bird's view, the name of each class is etched on the roof of each room. This immediately helps the users know which class they are looking at. In this view, hovering the cursor over each room shows a tool-tip text showing the name of the class in a larger font as shown in Figure \ref{fig:ClassNameToolTip}. The users can transition to the first-person view by clicking one of the code rooms and can go back to bird's view by pressing a keyboard shortcut. The camera's transition to and from bird's view is animated in order to preserve the user's sense of spatial awareness.

We improved bird's view mode further by adding syntax parsing features so as to be able to provide class-level details (such as the list of all member functions) to the user. In bird's view, right-clicking on a room shows a tool-tip balloon that provides a list of the methods defined in the class corresponding to the room (see Figure \ref{fig:MethodNameToolTip}). This allows the users to quickly glean useful information about each class. This overview is also available on one of the wallpapers inside each room as shown in Figure \ref{fig:WallMethod}.

By supporting syntax parsing, we implemented a commonly used feature in most IDEs, namely \textit{go-to definition}. This feature allows the programmers to quickly jump to the location inside a file where a user-defined type or a variable is defined for the first time. It aids the users in learning the codebase by allowing them to both mentally and visually connect the disparate parts of the code together.

In CP, the users can click on a variable, function or a user-defined type with their cross-hair and jump to the location where the clicked item is defined for the first time. The jump from a room to another room is done using two consecutive animated camera transitions: from the current room to bird's view and from bird's view to the room containing the definition\footnote{Unless the definition is inside the same room in which case the camera is transitioned to the definition directly.}. After the transition is finished, the definition of interest is signified by a blinking highlight to focus the user's attention and also to indicate the completion of the task. In addition to being visually appealing, these transitions maintain the users awareness of the 3D environment. It is worth mentioning that all movements and navigation actions in CP are done with animations instead of jumping around which helps preserve the spatial awareness of users. This way they can memorize location of each part of codebase such as classes or even methods and variables more efficiently.

%----------------------------------------------------------------------------------------
%	EVALUATION
%----------------------------------------------------------------------------------------
\section{EVALUATION: CodePark Usability}
%As mentioned in the Introduction, we designed CP to achieve three goals: ease of use, help with learning a codebase, and making code understanding fun and engaging. 
%
% To evaluate these goals and gauge the usability of our system, we designed and performed a user study. 
%As mentioned in Introduction our aforementioned research questions led directly to the development of CP and our hypotheses:
We conducted a usability study to evaluate the following hypotheses:

\begin{enumerate}
	\item[\textbf{H1:}] A project organized in a 3-space, city-like environment will be easier to learn.
	\item[\textbf{H2:}] A project organized in a game-like environment will be more engaging.
	\item[\textbf{H3:}] The move to 3-space will not make working with source more difficult.
\end{enumerate}	

At the time of this writing, CP only supports C\# projects. Therefore, we decided to compare CP with one of the most prominently used IDEs for C\#, namely Microsoft Visual Studio (VS). Given a few programming-related tasks, we are interested in studying the effects that using CP has and also determine how CP helps with code understanding.

%\subsection{Participants and Equipment}
\vspace{2mm}
\noindent
\textbf{Participants and Equipment.~}
We recruited 28 participants from University of Central Florida (22 males and 6 females ranging in age from 18 to 31 with a mean age of 22.8). Our requirements for participants were that they should be familiar with the C\# language and also have prior experience with VS. All participants were compensated with \$10 for their efforts. Each participant was given a pre-questionnaire containing some demographic questions as well as some questions asking about their experience in developing C\# applications. After that, a short C\# skill test was administered to validate their responses in the pre-questionnaire. The skill test contained five multiple-choice questions with varying difficulties selected from a pool of coding interview questions\footnote{\url{https://www.interviewmocha.com/tests/c-sharp-coding-test-basic}}. 

After the skill test, each participant was given a set of tasks to perform using each tool (VS or CP) and filled out post-questionnaires detailing their experience with each tool. Prior to performing the tasks on a specific tool, the participants were given a quick tutorial of both tools and were given a few minutes to warm up with the tool.

At the end of the study, the participants were asked to fill out a post-study questionnaire to share their overall experience. The duration of the user study ranged from 60 minutes to 90 minutes depending on how fast each participants progressed towards their assigned tasks.

Our setup consisted of a 50-inch Sony Bravia TV used as the computer screen. The users used Visual Studio 2013 Community Edition and Unity3D v5.4.0 on a machine running Microsoft\textsuperscript{\textregistered} Windows 10 64-bit equipped with 16.0 GB of RAM, Intel\textsuperscript{\textregistered} Core\textsuperscript{TM} i7-4790 processor with 4 cores running at 3.60 GHz and NVIDIA GeForce GTX 970 graphics processor.
\vspace{-1mm}

%\begin{figure}
% \centering
%   \includegraphics[width=0.8\columnwidth]{figures/Station.jpg}
%   \caption{The user study setup.}~\label{fig:Station}
%   \vspace{-5mm}
%\end{figure}

%\subsection{Experiment Design and Procedure}
\vspace{2mm}
\noindent
\textbf{Experiment Design and Procedure.~}
When comparing VS with CP, there are a few considerations involved in order to design a sound experiment that allows a fair comparison between the two tools. First, VS has been in development for many years and most C\# developers work with VS frequently. As a result, it could be the case that the users who use VS frequently are biased towards VS, because they have had more time to work and get comfortable with it. Consequently, devising a fair between-subject design would be difficult.

Second, focusing on a purely within-subject design presents other complications. It is important to avoid any unwanted learning effects in a within-subject design when the user is working with both tools. Given a particular codebase and a set of tasks, if a user performs those tasks in VS and then switches to CP to perform the same set of tasks, chances are that those tasks are performed much faster the second time. This is because the user will have learned the codebase's structure the first time and can leverage that knowledge in CP. To avoid this learning effect, the user should be given two different codebases to work with. However, care must be taken when selecting the two codebases, as they should be relatively similar in structure and the degree of difficulty. 

Having two codebases may pose another problem. Even if the two codebases are specifically chosen to be similar, minor differences between the two could affect the results. Moreover, studying and learning someone else's code can quickly become tedious and the users can become fatigued after using the first tool, affecting their performance in the second tool. Therefore, it is imperative to mitigate any of these unwanted effects in the study.

Facing with all these considerations, we opted to use a mixed-effects design for our study to benefit from both of the design modes. In our experiments, each participant used two different codebases with both tools. Several codebases were considered at first and after a few pilot runs and getting feedback from peers, two codebases were carefully selected. These selected codebases shared similar structures and properties. One codebase is a console-based library catalog system called Library Manager (LM). The other codebase is a console-based card matching game called Memory Game (MG). Table \ref{tab:codebases} summarizes these codebases. Note that even though MG contains more lines of code, some of its largest classes contain mostly duplicate code that handle drawing the shape of each playing card on the console window. Our assumption with the choice of codebases is that the two codebases are not significantly different and would not bias our results\footnote{This assumption will later be examined in the Discussion section.}.

To avoid the unwanted effects discussed previously, we permute the order of tools and codebases across participants. As a result, each participant started the experiment with either VS or CP. Also their first tasks were performed on either LM or MG. The possible permutations divided our participants into four groups detailed in Table \ref{tab:groups}. By recruiting 28 participants, we had 7 participants per group. We randomly assigned a participant to a group. This results in a balanced mixed-effects design in which all possibilities and orders are considered.

\begin{table}
\centering
\footnotesize
\setlength\tabcolsep{7pt}
\ra{1.2}
\caption[this is footnote]{Comparison of the two codebases used in the user studies: Library Manger (LM) and Memory Game (MG). \textit{LoC} is lines of code reported by the line counting program \textit{cloc}.\protect\footnotemark}
\begin{tabular}{ c c c c c}
	\toprule
	\textbf{Codebase} & \textbf{No. Classes} & \textbf{LoC} & \textbf{LoC (largest class)}\\
	\midrule
	LM & 14 & 977 & 237\\
	MG & 16 & 1753 & 791\\
	\bottomrule
\end{tabular}
\label{tab:codebases}
\end{table}
\footnotetext{\url{http://cloc.sourceforge.net/}}

\begin{table}
\centering
\footnotesize
\ra{1.1}
\setlength\tabcolsep{6.2pt}
\caption{Different experiment groups. Group names are only for tracking the number of participants in each permutation of the study.}
\begin{tabular}{ cc c c c c c}
	\toprule
	\textbf{Group} && \textbf{First} & \textbf{First} && \textbf{Second} & \textbf{Second} \\
	\textbf{Name} && \textbf{Tool} & \textbf{Codebase} && \textbf{Tool} & \textbf{Codebase} \\
	\midrule
	{A} && VS & LM && CP & MG\\
	{B} && CP & MG && VS & LM\\
	{C} && VS & MG && CP & LM\\
	{D} && CP & LM && VS & MG\\
	\bottomrule
\end{tabular}

%\vspace{-5mm}

\label{tab:groups}
\end{table}

On each codebase, the participants were asked to perform five tasks, each with a varying degree of difficulty. These tasks are presented in Table \ref{tab:tasks}. Among these, some tasks force the participant to explore the codebase, whereas other tasks were directly related to a participant's understanding of the codebase, the program structure and the logic behind it. Tasks T1 and T2 were similar for both codebases. Task T3 asked about the object-oriented relationship between classes \textit{A} and \textit{B}. In LM, this relationship was inheritance and in MG this relationship was having a common parent class (\textit{A} and \textit{B} are siblings)\footnote{The selected classes were the same for all participants.}. Prior to being tasked with T4, the participants were asked to try out a particular feature of the program. Upon the trial, the program would crash prematurely and no output was produced. The participants were told that the reason for this behavior was a simple intentional bug and were tasked with finding the likely location of the bug. For task T5, the participants were asked to imagine a scenario where somebody asked them about their approach for adding a new feature to the particular codebase that they were working on. In LM, they were asked to add additional menu options for specific types of users. In MG, they were asked to modify the scoring system to penalize the player for each mistake. We should note that neither of these tasks involve writing any code. This was necessary because at the time of this writing, CP did not incorporate a code editor. Also, we are primarily interested in determining the effects of CP on code understanding. The participants were responsible for showing a suitable location in the logic to add these features. There were multiple valid locations for each task in either codebase and the participants were free to select any of the valid locations. When performing any of these tasks, the users were explicitly told that they were not allowed to use the debugging features of VS, nor the search functionality for finding a specific type or class. They were, however, permitted to use VS's built-in \textit{go-to definition} feature by holding down the control key and clicking with the mouse on a user-defined type. After performing the tasks, the participants were given three questionnaires to fill out. 

%\subsection{Metrics}
\vspace{2mm}
\noindent
\textbf{Metrics.~}
For quantitative data, we recorded the time each participant took to perform each task. The qualitative measurement was performed using the post-task and post-study questionnaires. After performing the tasks with each tool, the participants were given a post-task questionnaire that asked them to share their experience about the tool they worked with. These questions were the same for both tools and are shown in Table \ref{tab:postTaskQuestionnaire}. The responses to these questions were measured on a 7-point Likert scale (1 = the most negative response, 7 = the most positive response).

Upon the completion of all tasks, the participants were given a post-study questionnaire to measure their preferences of both tools from different aspects. This questionnaire is detailed in Table \ref{tab:postOverallQuestionnare}. The participants were required to select either VS or CP in their responses to each question.

\begin{table}
\centering
\footnotesize
\setlength\tabcolsep{7pt}
\ra{1.2}
\caption{Participant tasks. Each task was timed. We used these measurements for our quantitative analysis.}
\begin{tabular}{l l }
	\toprule
	\multicolumn{2}{c}{\textbf{Task}} \\
	\midrule
	\textbf{T1}& Find a valid username to login into the program.\\
	\textbf{T2}& Find an abstract class in the codebase.\\
	\textbf{T3}& Determine the relationship between classes \textit{A} and \textit{B}.\\
	\textbf{T4}& Find an intentional bug that causes a program crash.\\
	\textbf{T5}& Pinpoint a reasonable location in the code for adding the\\
	           & necessary logic to support feature \textit{X}.\\
	\bottomrule
\end{tabular}

%\vspace{-5mm}
\label{tab:tasks}
\end{table}

\begin{table}
\centering
\footnotesize
\setlength\tabcolsep{1.5pt}
\ra{1.2}
\caption{Post-task questionnaire. Participants answered
these questions on a 7-point Likert scale after finishing their tasks with both tools. We used this data for our qualitative analysis.}
\begin{tabular}{ l l }
	\toprule
	\multicolumn{2}{c}{\textbf{Post Task Questionnaire}} \\
	\midrule
	\textbf{Q1}& I found it easy to work with CP/VS.\\
	\textbf{Q2}& I found it easy to become familiar with CP/VS.\\
	\textbf{Q3}& CP/VS helps me become familiar with codebase's structure.\\
	\textbf{Q4}& It was easy to navigate through the code with CP/VS.\\
	\textbf{Q5}& It was easy to find the definition of some variable with CP/VS.\\
	\textbf{Q6}& How much did you like CP/VS?\\
	\textbf{Q7}& How did you feel when using the tool?\\
	\textbf{Q8}& It was easy to find what I wanted in the code using CP/VS.\\
	\bottomrule
\end{tabular}
\label{tab:postTaskQuestionnaire}
\end{table}

\begin{table}[t]
\centering
\footnotesize
\ra{1.2}
\caption{Post-study questionnaire. Participants answered
these questions after finishing all their tasks with both tools. The answer to each question is either CP or VS. We used this data for qualitative analysis.}
\begin{tabular}{ l l }
	\toprule
	\multicolumn{2}{c}{\textbf{Post Study Questionnaire}} \\
	\midrule
	\textbf{SQ1} & 	Which tool is more comfortable to use?\\
	\textbf{SQ2} &	Which tool is more likable?\\
	\textbf{SQ3} &	Which tool is more natural? \\
	\textbf{SQ4} &	Which tool is easier to use?\\
	\textbf{SQ5} & 	Which tool is more fun to use?\\
	\textbf{SQ6} &	Which tool is more frustrating? \\
	\textbf{SQ7} & 	Which tool helps you more in remembering the\\
                 &  codebase structure?\\
	\textbf{SQ8} &	Which tool do you prefer for learning a codebase?\\
	\textbf{SQ9} &	Which tool do you prefer for a codebase\\
		         & you are already familiar with for additional work? \\
	\textbf{SQ10} &	Which tool do you prefer for finding a particular\\
	              & class/variable?\\
	\textbf{SQ11} &	Which tool do you prefer for tracking down a bug?\\
	\textbf{SQ12} &	Overall, which tool is better? \\
\bottomrule
\end{tabular}
%\vspace{-5mm}
\label{tab:postOverallQuestionnare}
\end{table}

\subsection{Results}
As mentioned before, we recorded quantitative as well as qualitative data. To analyze these data, we divided all of our participants into two equally sized groups based on their experience in C\# and software development. We leveraged the results of the C\# skill-test as well as the self-declared answers to determine the skill level of each participants. The skill-test questions were weighted twice as much in order to reduce potential over- or undervaluation of the self-declared responses.

In summary, our experiments have three factors: tool, codebase and experience. The tool factor has two levels: CP or VS, the codebase factor has two levels: LM or MG and the experience factor has two levels: beginner or expert.

\subsubsection{Quantitative Results}

%=======================================================================
%
%                              CHARTS
%
%=======================================================================
\begin{figure*}[t]
\centering
\begin{subfigure}{0.3\textwidth}
  \includegraphics[width=\columnwidth]{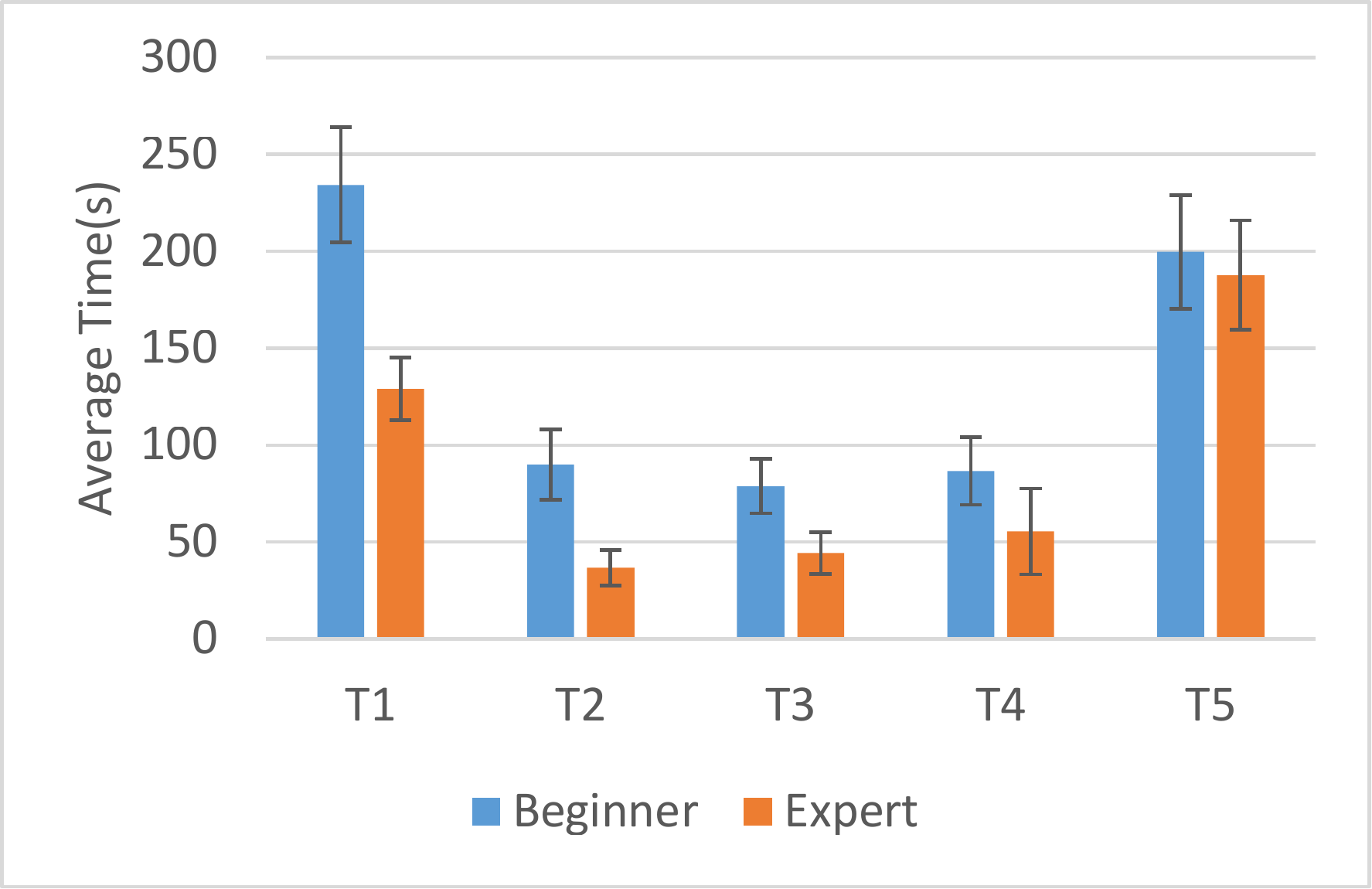}
\caption{MTTC by experience level.} 
\label{fig:ChartTaskTimeExperience}
\end{subfigure}
%\hspace*{\fill} % separation between the subfigures
\begin{subfigure}{0.3\textwidth}
  \includegraphics[width=\columnwidth]{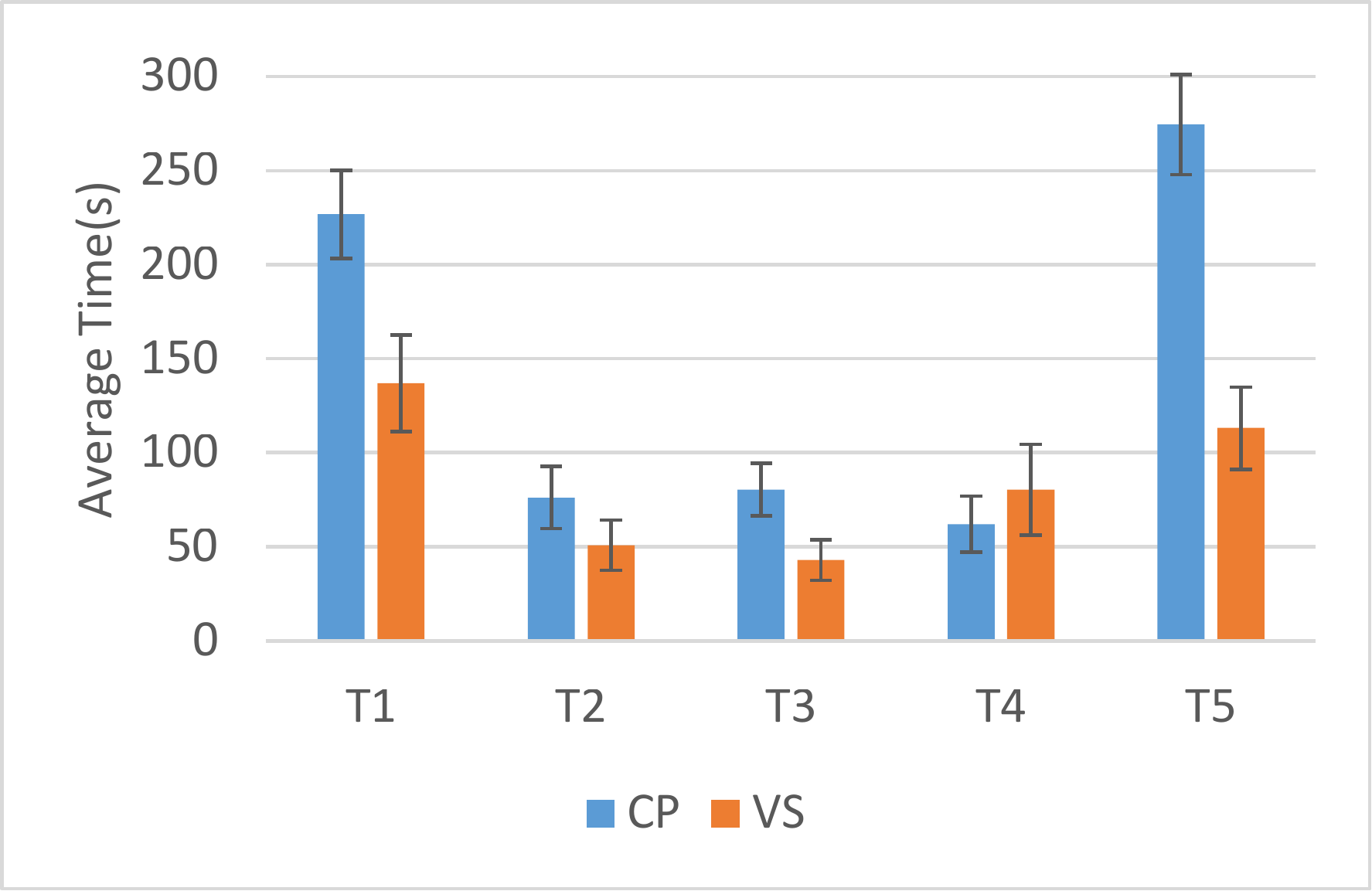}
\caption{MTTC based on the tool used.} \label{fig:ChartTaskTimeTool}
\end{subfigure}
%\hspace*{\fill} % separation between the subfigures
\begin{subfigure}{0.3\textwidth}
  \includegraphics[width=\columnwidth]{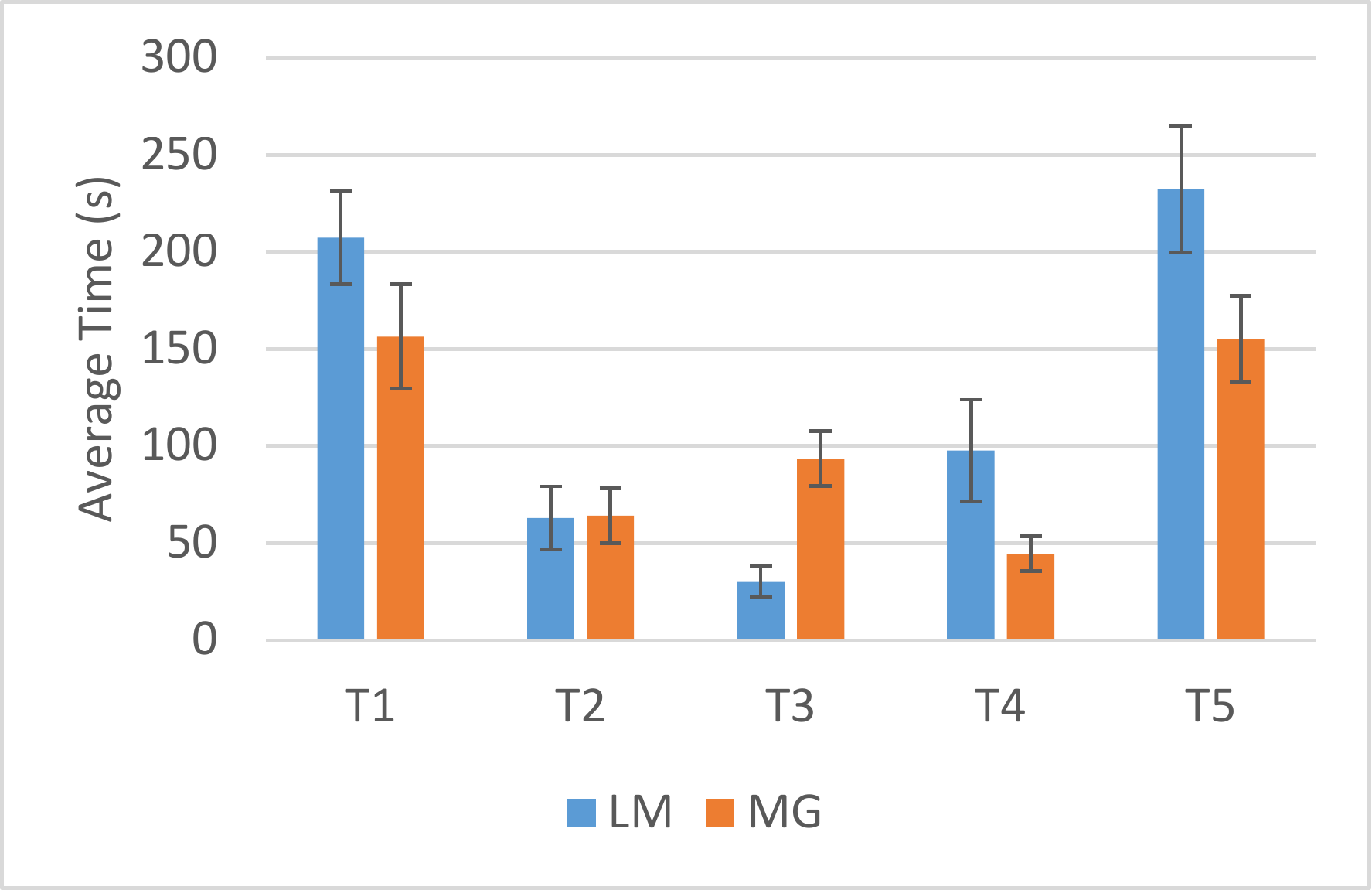}
\caption{MTTC based on the codebase.} \label{fig:ChartTastTimeCodeBase}
\end{subfigure}

\caption{Mean time to task completion (MTTC). All time values are reported in seconds.} 
\label{fig:charts}
\end{figure*}

%=======================================================================
%
%                              TABLE ART
%
%=======================================================================
\begin{table*}[t]
\centering
\scriptsize
\setlength\tabcolsep{4.8pt}
\ra{1.2}
\caption{ANOVA results on the quantitative data (time to task completion). Statistically significant results (95\% confidence) are highlighted in gray. The first column (TT) is task time. Codebase is abbreviated as \textit{CB}. Experience level is abbreviated as \textit{Exp.} }
\begin{tabular}{c c c c c c c c c c c c c c c c c c c c c c}
	\toprule
	\multirow{2}{*}{\textbf{TT}} 
	&& \multicolumn{2}{c}{\textbf{CB}} 
	&& \multicolumn{2}{c}{\textbf{Tool}} 
	&& \multicolumn{2}{c}{\textbf{Exp.}}
	&& \multicolumn{2}{c}{\textbf{Tool$\times$CB}}  
	&& \multicolumn{2}{c}{\textbf{CB$\times$Exp.}} 
	&& \multicolumn{2}{c}{\textbf{Tool$\times$Exp.}}
	&& \multicolumn{2}{c}{\textbf{ Tool$\times$CB$\times$Exp.}} 
	 \\
	 \cmidrule{3-4}  \cmidrule{6-7} \cmidrule{9-10} \cmidrule{12-13}  \cmidrule{15-16} \cmidrule{18-19} \cmidrule{21-22} 
	 && \multicolumn{1}{l}{$F_{1,24}$} & \multicolumn{1}{r}{$p$-value}
	 && \multicolumn{1}{l}{$F_{1,24}$} & \multicolumn{1}{r}{$p$-value}
	 && \multicolumn{1}{l}{$F_{1,24}$} & \multicolumn{1}{r}{$p$-value}
	 && \multicolumn{1}{l}{$F_{1,24}$} & \multicolumn{1}{r}{$p$-value}
	 && \multicolumn{1}{l}{$F_{1,24}$} & \multicolumn{1}{r}{$p$-value}
	 && \multicolumn{1}{l}{$F_{1,24}$} & \multicolumn{1}{r}{$p$-value}
	 &&  \multicolumn{1}{l}{$F_{1,24}$} & \multicolumn{1}{r}{$p$-value}
	 \\
	\midrule
	\textbf{T1}
	&&$3.92$ & $=0.06$
	&&\cellcolor{gray!25} $12.0$ & \cellcolor{gray!25} $<0.05$
	&&\cellcolor{gray!25}$7.53$ & \cellcolor{gray!25}$<0.05$
	&&$0.78$ & $=0.39$
	&&$0.49$ & $=0.49$
	&&$0.19$ & $=0.66$
	&&$0.10$ & $=0.75$
	\\
	\textbf{T2}
	&&$1.18$ & $=0.29$
	&&$0.32$ & $=0.58$
	&&$3.71$ & $=0.07$
	&&$0.46$ & $=0.50$
	&&$0.63$ & $=0.44$
	&&$0.04$ & $=0.84$
	&&$0.46$ & $=0.51$
	\\
	\textbf{T3}
	&&\cellcolor{gray!25}$45.2$ & \cellcolor{gray!25}$<0.05$
	&&\cellcolor{gray!25}$15.9$ & \cellcolor{gray!25}$<0.05$
	&&\cellcolor{gray!25}$5.70$ & \cellcolor{gray!25}$<0.05$
	&&$1.02$ & $=0.32$
	&&$0.58$ & $=0.46$
	&&\cellcolor{gray!25}$6.01$ & \cellcolor{gray!25}$<0.05$
	&&$0.00$ & $=0.95$
	\\
	\textbf{T4}
	&&$0.70$ & $=0.41$
	&&$0.03$ & $=0.86$
	&&$3.56$ & $=0.07$
	&&$0.15$ & $=0.70$
	&&$0.11$ & $=0.74$
	&&$0.10$ & $=0.76$
	&&$0.71$ & $=0.41$
	\\
	\textbf{T5}
	&&\cellcolor{gray!25}$7.71$ & \cellcolor{gray!25}$<0.05$
	&&\cellcolor{gray!25}$41.9$ & \cellcolor{gray!25}$<0.05$
	&&$0.08$ & $=0.78$
	&&$3.93$ & $=0.06$
	&&$1.35$ & $=0.26$
	&&$0.32$ & $=0.58$
	&&$0.05$ & $=0.82$
	\\
%	\textbf{Total}
%	&&$5.03$ & $<0.05$
%	&&$55.0$ & $<0.05$
%	&&$7.16$ & $<0.05$
%	&&$0.03$ & $=0.87$
%	&&$1.00$ & $=0.33$
%	&&$0.22$ & $=0.64$
%	&&$0.01$ & $=0.92$
%	\\

\bottomrule
\end{tabular}

\label{tab:task-time}
\end{table*}

Our quantitative results are based on the time a participant took to complete an assigned task. When validating ANOVA assumptions, we found that most group response variables failed the Shapiro-Wilk normality tests. Since our factorial design contains 3 factors (Codebase$\times$Tool$\times$Experience) with two levels each, we decided to utilize the Aligned Rank Transform (ART) \cite{wobbrock2011art} to make the data suitable for ANOVA. The mean time task completion results could be find in Figure \ref{fig:charts}\footnote{All error bars are standard error values.}. The analysis of these results are presented in Table \ref{tab:task-time}.

\subsubsection{Qualitative Results}
Our qualitative results are comprised of two parts. The first part consists of the responses of the participants to the Likert-scale questions in the post-task questionnaire. The second part consists of the responses of the participants to the questions in the post-study questionnaire.

\paragraph{Post-Task Questionnaire}

\begin{table*}[t]
\centering
\scriptsize
\setlength\tabcolsep{4.3pt}
\ra{1.2}
\caption{ANOVA results on the qualitative data (post-task questionnaire). Statistically significant results (95\% confidence) are highlighted in gray. Codebase is abbreviated as \textit{CB}. Experience level is abbreviated as \textit{Exp.} }
\begin{tabular}{c c c c c c c c c c c c c c c c c c c c c c}
	\toprule
	\multirow{2}{*}{\textbf{Question}} 
	&& \multicolumn{2}{c}{\textbf{CB}} 
	&& \multicolumn{2}{c}{\textbf{Tool}} 
	&& \multicolumn{2}{c}{\textbf{Exp.}}
	&& \multicolumn{2}{c}{\textbf{Tool$\times$CB}}  
	&& \multicolumn{2}{c}{\textbf{CB$\times$Exp.}} 
	&& \multicolumn{2}{c}{\textbf{Tool$\times$Exp.}}
	&& \multicolumn{2}{c}{\textbf{ Tool$\times$CB$\times$Exp.}} 
	 \\
	 \cmidrule{3-4}  \cmidrule{6-7} \cmidrule{9-10} \cmidrule{12-13}  \cmidrule{15-16} \cmidrule{18-19} \cmidrule{21-22} 
	 && \multicolumn{1}{l}{$F_{1,24}$} & \multicolumn{1}{r}{$p$-value}
	 && \multicolumn{1}{l}{$F_{1,24}$} & \multicolumn{1}{r}{$p$-value}
	 && \multicolumn{1}{l}{$F_{1,24}$} & \multicolumn{1}{r}{$p$-value}
	 && \multicolumn{1}{l}{$F_{1,24}$} & \multicolumn{1}{r}{$p$-value}
	 && \multicolumn{1}{l}{$F_{1,24}$} & \multicolumn{1}{r}{$p$-value}
	 && \multicolumn{1}{l}{$F_{1,24}$} & \multicolumn{1}{r}{$p$-value}
	 &&  \multicolumn{1}{l}{$F_{1,24}$} & \multicolumn{1}{r}{$p$-value}
	 \\
	\midrule
	\textbf{Q1}
	&&$1.00$ & $=0.33$
	&&$0.94$ & $=0.34$
	&&$1.94$ & $=0.18$
	&&$2.35$ & $=0.14$
	&&$0.55$ & $=0.47$
	&&\cellcolor{gray!25}$9.10$ &\cellcolor{gray!25} $<0.05$
	&&$0.06$ & $=0.81$
	\\
	\textbf{Q2}
	&&$0.91$ & $=0.35$
	&&\cellcolor{gray!25}$7.98$ &\cellcolor{gray!25} $<0.05$
	&&\cellcolor{gray!25}$7.81$ &\cellcolor{gray!25} $<0.05$
	&&$2.45$ & $=0.13$
	&&$0.61$ & $=0.44$
	&&$0.00$ & $=0.98$
	&&$0.00$ & $=0.98$
	\\
	\textbf{Q3}
	&&$0.02$ & $=0.89$
	&&\cellcolor{gray!25}$14.2$ &\cellcolor{gray!25} $<0.05$
	&&$2.35$ & $=0.14$
	&&$0.70$ & $=0.41$
	&&$0.57$ & $=0.46$
	&&$0.04$ & $=0.84$
	&&$0.08$ & $=0.77$
	\\
	\textbf{Q4}
	&&$0.10$ & $=0.75$
	&&$2.10$ & $=0.16$
	&&\cellcolor{gray!25}$8.18$ &\cellcolor{gray!25} $<0.05$
	&&$0.34$ & $=0.56$
	&&$0.12$ & $=0.74$
	&&$0.81$ & $=0.38$
	&&$0.38$ & $=0.54$
	\\
	\textbf{Q5}
	&&$0.97$ & $=0.33$
	&&$0.16$ & $=0.69$
	&&$1.07$ & $=0.31$
	&&\cellcolor{gray!25}$4.44$ &\cellcolor{gray!25} $<0.05$
	&&$0.02$ & $=0.90$
	&&$0.23$ & $=0.64$
	&&$0.16$ & $=0.69$
	\\
	\textbf{Q6}
	&&$0.38$ & $=0.54$
	&&$0.28$ & $=0.60$
	&&$3.35$ & $=0.08$
	&&$0.14$ & $=0.71$
	&&$0.00$ & $=0.97$
	&&$0.09$ & $=0.77$
	&&$0.07$ & $=0.80$
	\\
	\textbf{Q7}
	&&$0.09$ & $=0.77$
	&&$1.10$ & $=0.30$
	&&\cellcolor{gray!25}$11.3$ &\cellcolor{gray!25} $<0.05$
	&&$0.43$ & $=0.52$
	&&$0.05$ & $=0.82$
	&&$0.10$ & $=0.75$
	&&$0.28$ & $=0.60$
	\\
	\textbf{Q8}
	&&$0.11$ & $=0.74$
	&&$0.27$ & $=0.61$
	&&$3.04$ & $=0.09$
	&&$0.27$ & $=0.61$
	&&$0.67$ & $=0.42$
	&&$0.03$ & $=0.86$
	&&$0.00$ & $=0.95$
	\\

\bottomrule
\end{tabular}

\label{tab:PostTaskQuestionnareARTAnalyze}
\end{table*}

\begin{figure}
\centering
  \includegraphics[width=0.9\columnwidth]{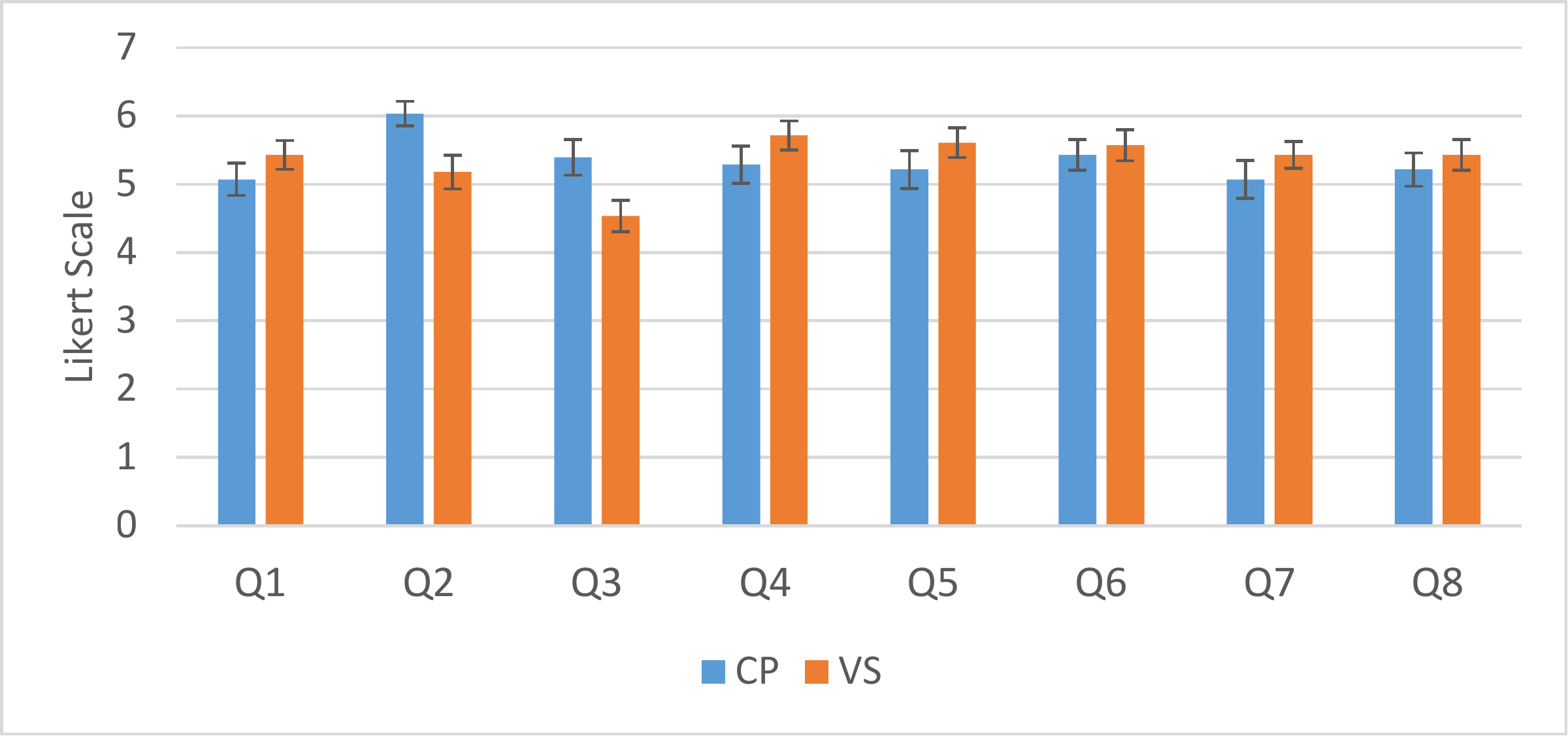}
\caption{Mean responses to the post-study questionnaire for CP and VS.} 
\label{fig:ChartPostTask}
\vspace{-5mm}
\end{figure}

The responses to the Likert-scale questions in our post-task questionnaire failed the Shapiro-Wilk normality tests. As such, similar to our quantitative results, we utilized ART \cite{wobbrock2011art} for ANOVA. Table \ref{tab:PostTaskQuestionnareARTAnalyze} summarizes the analyses of our data. Average ratings for the eight post-task questions are shown in Figure \ref{fig:ChartPostTask}.

%Our tests were not powerful enough to detect significant main effects caused by the codebase on the participants' responses. However, the tool had a significant difference on the responses to the questions, namely Q2 and Q3. Results show that the participants found it easier to become familiar with CP compared to VS \anova{7.98}{0.01}. They also found CP to be more helpful in becoming familiar with codebase's structure compared to VS \anova{14.21}{0.01}. The significant differences based on experience is not related to our study. Also there was an interaction effect between tool and experience on how much a tool was easy to work with (Q1) ($F_{1,24} = 9.10,~p < 0.05,~\eta^{2}_{p} = 0.27~\text{(L)}$). There was also another interaction effect between tool and codebase on the responses to Q5 which measured the ease of finding definition ($F_{1,24} = 4.44,~p < 0.05,~\eta^{2}_{p} = 0.37~\text{(L)}$).

\paragraph{Post-Study Questionnaire}

The post-study questionnaire required the participant to pick one tool for each question (see Table \ref{tab:postOverallQuestionnare}). Since each question had only two choices, we used Chi-squared test to analyze the results. Table \ref{tab:postOverallQuestionnareAnalysis} summarizes the analyses of our data. We noticed that a few participants left some of the questions unanswered.

%The results indicate that there was a significant difference in tool comfort (SQ1) \chisq{28}{7.00}. Also CP was significantly more likable than VS  (SQ2) \chisq{28}{5.14}. Every single participant believed that CP was more fun to use compared to VS (SQ5) \chisq{28}{28.00}. CP also helped the participants in remembering codebase's structure significantly more than VS (SQ7) \chisq{28}{11.57}. For learning a codebase the participants preferred CP significantly more than VS (SQ8) \chisq{28}{7.00}. In other questions there was no significant deference between CP and VS.

\begin{table}
\centering
\scriptsize
\setlength\tabcolsep{5.3pt}
\ra{1.2}
\caption{Chi-squared analysis on the post-study responses. Statistically significant results (95\% confidence) are highlighted. The numbers in VS and CP columns represent the total times each tool was selected by the participant in response to a question. The winner in each significant category is highlighted.}
\begin{tabular}{cc c c c c l l}
	\toprule
	\textbf{Question}&& \textbf{VS} && \textbf{CP} && \multicolumn{2}{c}{\textbf{Chi-squared Test}}  \\
	\midrule
	\textbf{SQ1} && \cellcolor{gray!25}21 && 7 &&\cellcolor{gray!25}$\chi^2(1,~N=28)~=7.00$ &\cellcolor{gray!25}$p<0.05$ \\
	\textbf{SQ2} && 8 && \cellcolor{gray!25}20 && \cellcolor{gray!25}$\chi^2(1,~N=28)~=5.14$ &\cellcolor{gray!25}$p<0.05$ \\
	\textbf{SQ3} && 16 && 12 && $\chi^2(1,~N=28)~=0.57$ & $p=0.45$ \\
	\textbf{SQ4} && 17 && 11 && $\chi^2(1,~N=28)~=1.29$ & $p=0.26$ \\
	\textbf{SQ5} && 0 && \cellcolor{gray!25}28 &&\cellcolor{gray!25}$\chi^2(1,~N=28)~=28.00$ &\cellcolor{gray!25}$p<0.05$ \\
	\textbf{SQ6} && 11 && 13 && $\chi^2(1,~N=24)~=0.17$ & $p=0.68$ \\
	\textbf{SQ7} && 5 && \cellcolor{gray!25}23 && \cellcolor{gray!25}$\chi^2(1,~N=28)~=11.57$ & \cellcolor{gray!25}$p<0.05$ \\
	\textbf{SQ8} && 7 && \cellcolor{gray!25}21 &&\cellcolor{gray!25}$\chi^2(1,~N=28)~=7.00$ &\cellcolor{gray!25}$p<0.05$ \\
	\textbf{SQ9} && 19 && 9 && $\chi^2(1,~N=28)~=3.57$ & $p=0.06$ \\
	\textbf{SQ10} && 15 && 13 && $\chi^2(1,~N=28)~=0.14$ & $p=0.71$ \\
	\textbf{SQ11} && 19 && 9 && $\chi^2(1,~N=28)~=3.57$ & $p=0.06$ \\
	\textbf{SQ12} && 16 && 11 && $\chi^2(1,~N=27)~=0.93$ & $p=0.34$ \\
\bottomrule
\end{tabular}
\label{tab:postOverallQuestionnareAnalysis}
\end{table}

\subsection{Discussion}
The goal of this user study was to determine the degree to which we achieved our design goals with CP. Specifically, we were interested in determining how much it helps in learning a codebase (H1), how engaging is to work with CP especially for novice users (H2), how the users feel about working with it and how it compares against a traditional IDE such as VS (H3). Our results can be discussed from various aspects.

On the tool level and from a qualitative aspect, it is evident that the participants found CP significantly easier than VS to get familiar with, even though all participants had prior familiarity and experience with VS. The participants also found CP to be significantly more beneficial in becoming familiar with a codebase compared to VS. Both of these results were obtained regardless of the participant's experience, \textit{i.e.} both experts and beginners found CP to be superior than VS in these two categories and this confirm our first and third hypotheses. Referring to the post-study questionnaire (see Table \ref{tab:postOverallQuestionnare}), we see that the participants believed CP to be more likable compared to VS. Further, they believed that CP not only helped in learning the code structure but also helped them in remembering the code they studied and also finding the code elements that they were looking for. The interesting observation here is that all 28 participants unanimously believed that CP was more enjoyable which indicates that we achieved our goal of designing an engaging tool (H2). Note that the results in all these categories were statistically significant. These results are further corroborated by the written feedback that our participants provided. Participants found CP to be \textit{``generally easier [than VS] to understand the structure of the code"} and also felt that they were \textit{``using [their] spatial memory"} when working with CP. These results show that regardless of the participant's experience our three hypotheses are valid.

In the remaining qualitative categories, there was no significant difference observed in users' responses between CP and VS. This is again interesting because VS has been in development since 1997 and has gone through many iterations and refinements, whereas CP has only been in development for about 6 months.

\begin{figure}
\centering
\begin{subfigure}{0.4\columnwidth}
  \includegraphics[width=\columnwidth]{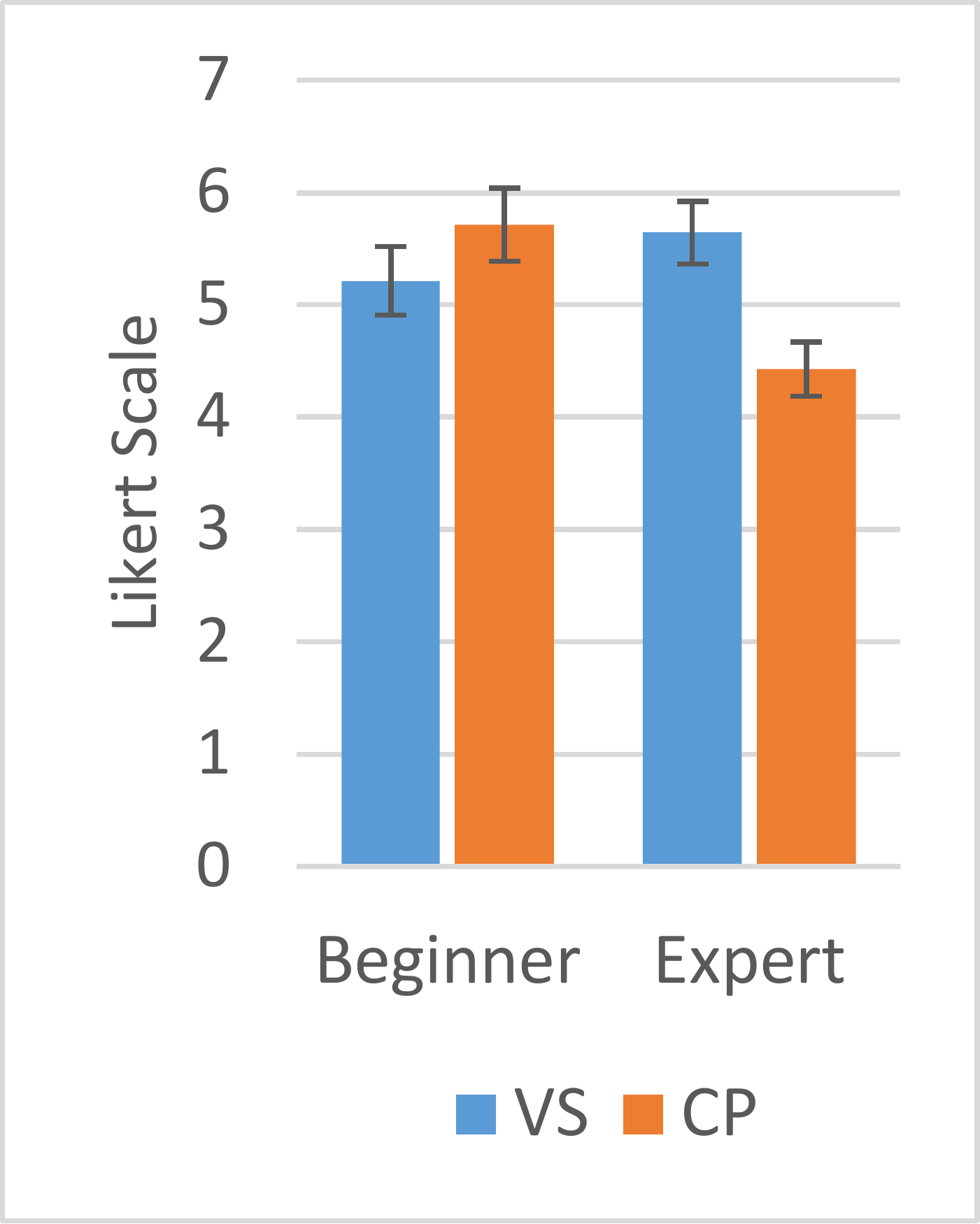}
\caption{Q1} \label{fig:ChartInteractionTool_LVLEdited}
\end{subfigure}
%\hspace*{\fill} % separation between the subfigures
\begin{subfigure}{0.4\columnwidth}
  \includegraphics[width=\columnwidth]{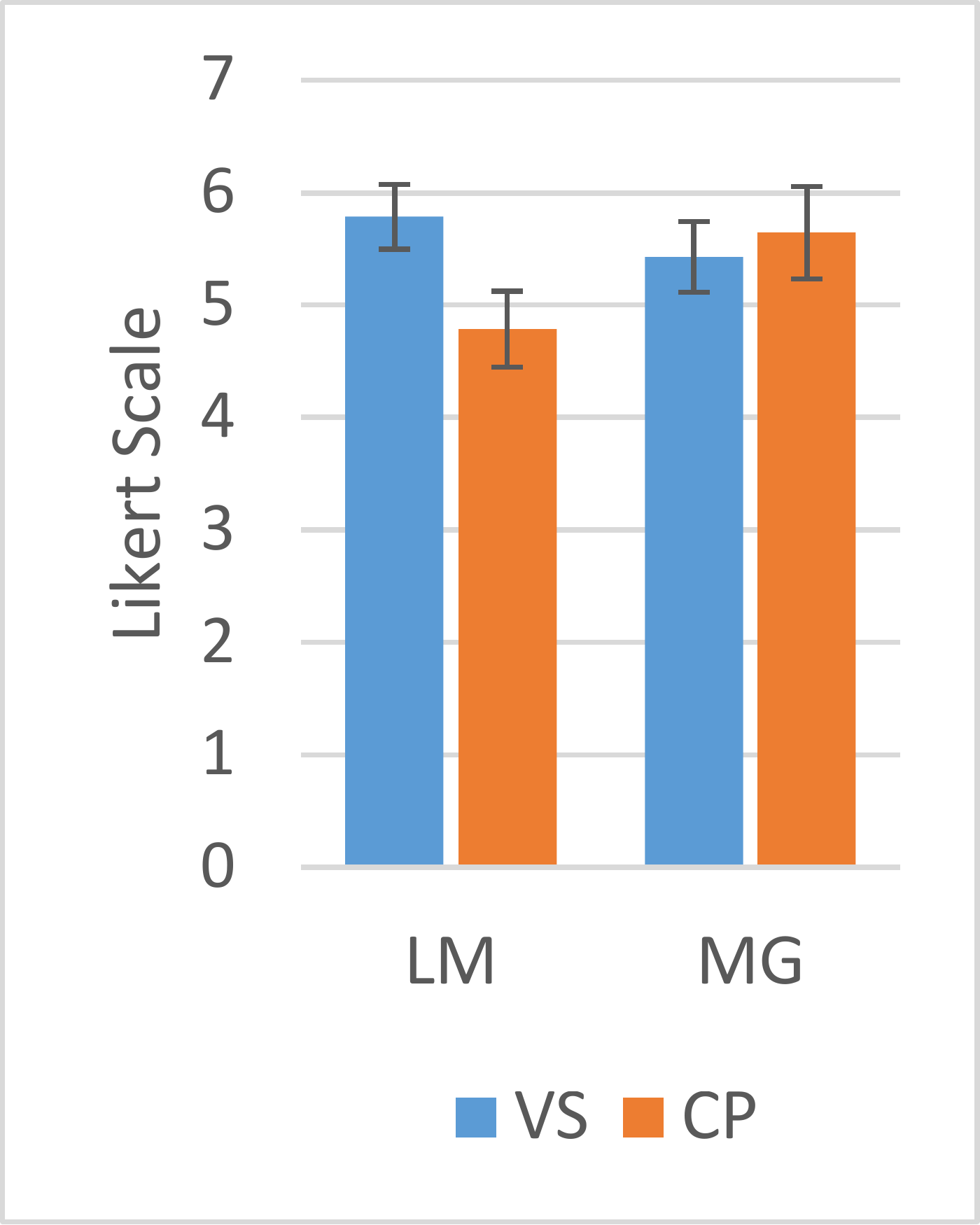}
\caption{Q5} \label{fig:ChartInteractionTool_CBEdited}
\end{subfigure}

\caption{(a) Difference of averages between VS and CP grouped by user experience for Q1 responses. (b) Difference of averages between VS and CP grouped by codebase for Q5 responses. } 
\label{fig:InteractionCharts}
%\vspace{-5mm}
\end{figure}

Our results indicate an interaction between the tool's ease of use and the participants' experience levels (Q1). A closer look at the responses (see Figure \ref{fig:ChartInteractionTool_LVLEdited}) reveals that the beginners found CP to be easier to work with. Conversely, experts found VS to be easier. This coincides with what one would expect: the prior experience of the experts with VS gives them an edge in performing the assigned task. As a result, those participants found VS easier to work with compared to CP. When asked about their opinion, one user described CP as \textit{``an amusement park for beginners"}. Another user told us: \textit{``With more polish (smoother experience), I think there is promise for CP (especially for new programmers)"}.

We observe an interaction effect between tool and the codebase for users' perceived facility with finding variable definitions (Q5). The mean responses for this question are detailed for each codebase in Figure \ref{fig:ChartInteractionTool_CBEdited}. From LM to MG, we see an increase in the mean response values for CP, but a decrease in the responses for VS. One possible explanation for the small superiority  of CP over to VS on MG is that MG contains more code with larger classes (see Table \ref{tab:codebases}). It could be that this larger size and the existence of more clutter give CP a slight edge when finding the definition of variables. This could potentially lead to the conclusion that CP is a more favorable tool for larger projects. However, this conclusion is premature and we believe a more detailed study is warranted to examine this interaction in more details.

On the tool level and from a quantitative point of view, it is evident that the choice of tool had a significant effect on time to completion of three tasks. Finding a valid username to login into the program (T1), determining the relationship between two classes (T3) and pinpointing a reasonable location in the code for adding the necessary logic to support some feature (T5). In all these cases, participants took significantly less time to complete their tasks with VS compared to CP. There are several possible explanations for this observations. 

The first is the existence of transition animations in CP. As discussed in \nameref{sec:ui}, the animations were necessary to preserve the user's sense of environmental awareness. At any instance of time, the length of these animations depend on the camera's relative position to each code room and also the size of the codebase. Nevertheless, every animation was at least 1.5 seconds long. Considering a hypothetical situation where a user is in bird's view and wants to switch to first-person mode to inspect a code room and then go back to bird's view, the transition animations take about 3 seconds. It takes a significantly less amount of time to open two files consecutively in VS. 

\begin{figure}
\centering
  \includegraphics[width=0.55\columnwidth]{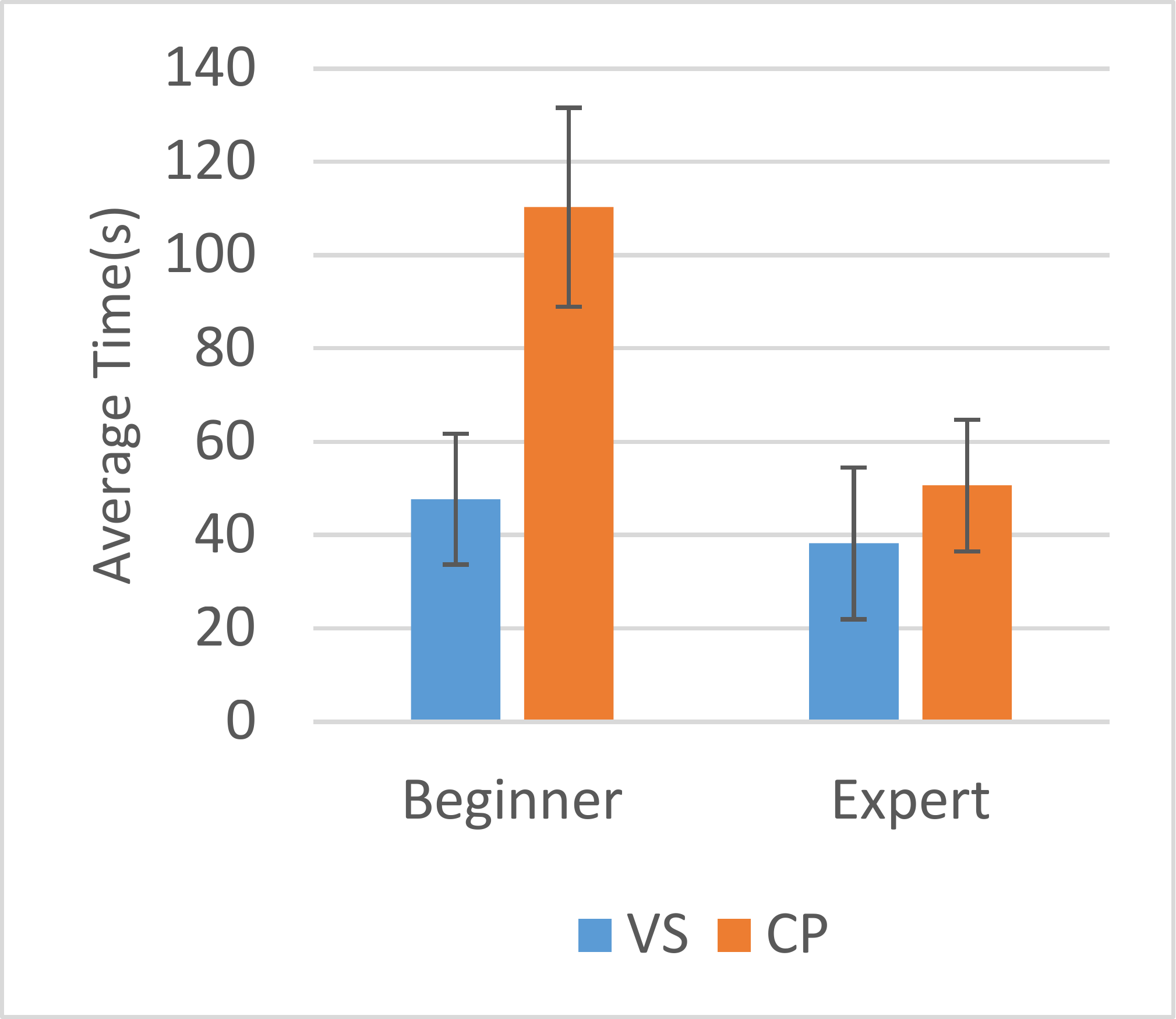}
\caption{The average time spent on completing T3 based on each tool and the participants' experience level.} 
\label{fig:ChartInteractionTaskTimeTool_LVLEdited}
%\vspace{-3pt}
\end{figure}

\begin{figure*}[ht]
\centering
\begin{subfigure}{0.3\textwidth}
  \includegraphics[width=\columnwidth]{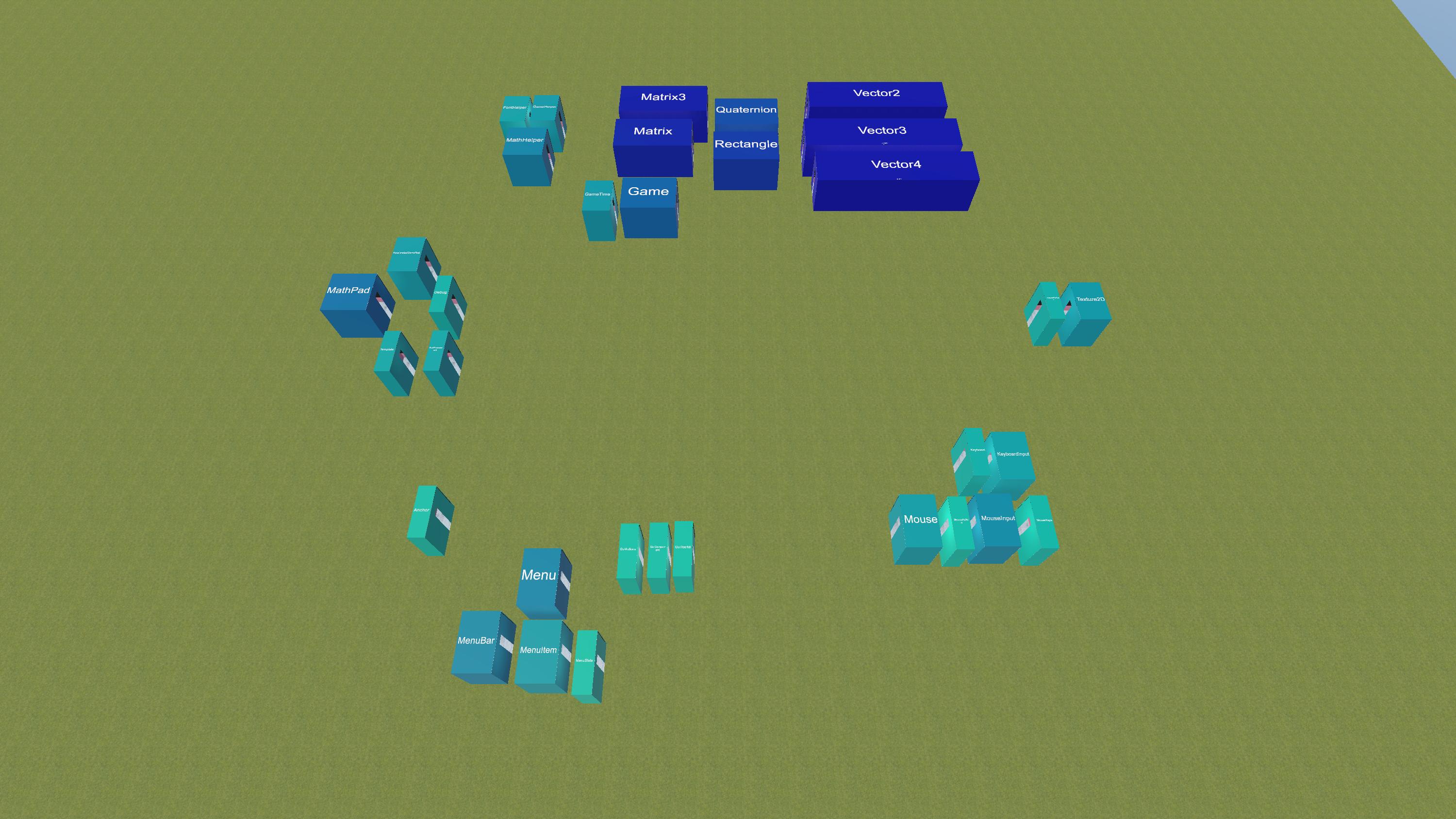}
  \caption{Participant 1} \label{fig:screenGame1}
\end{subfigure}
\begin{subfigure}{0.3\textwidth}
  \includegraphics[width=\columnwidth]{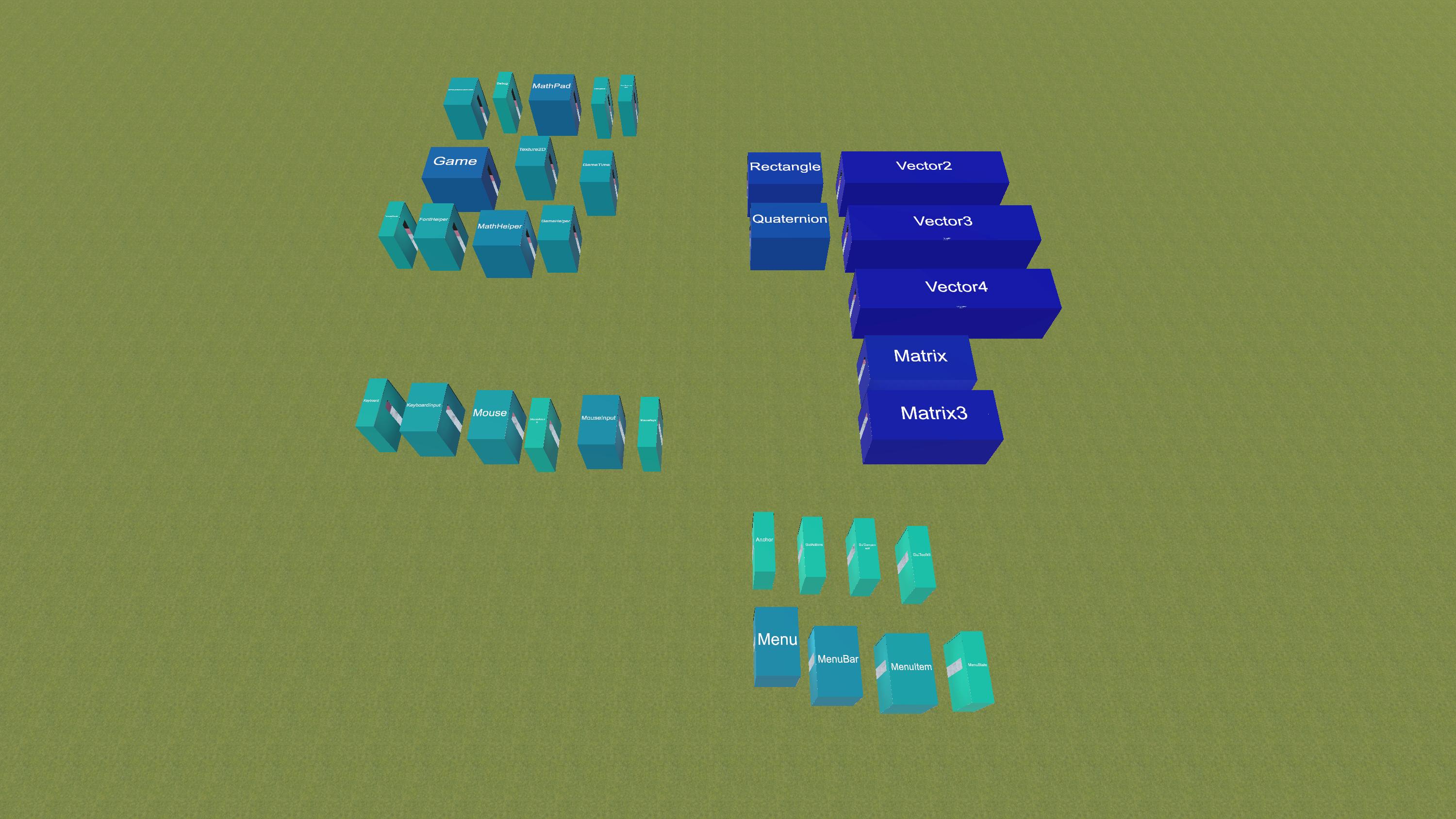}
  \caption{Participant 2} \label{fig:screenGame2}
\end{subfigure}
\begin{subfigure}{0.3\textwidth}
  \includegraphics[width=\columnwidth]{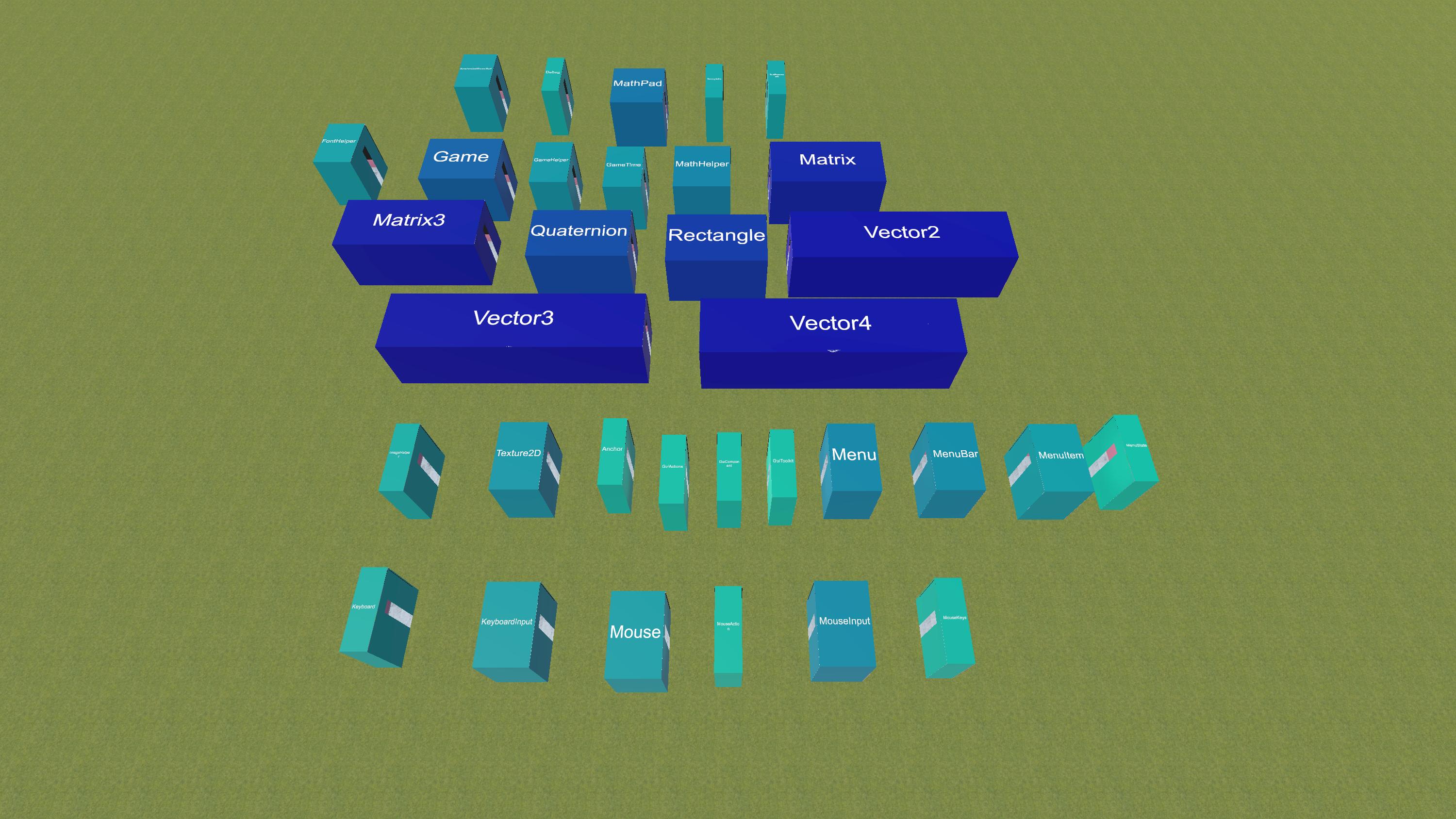}
  \caption{Participant 3} \label{fig:screenGame3}
\end{subfigure}
\begin{subfigure}{0.3\textwidth}
  \includegraphics[width=\columnwidth]{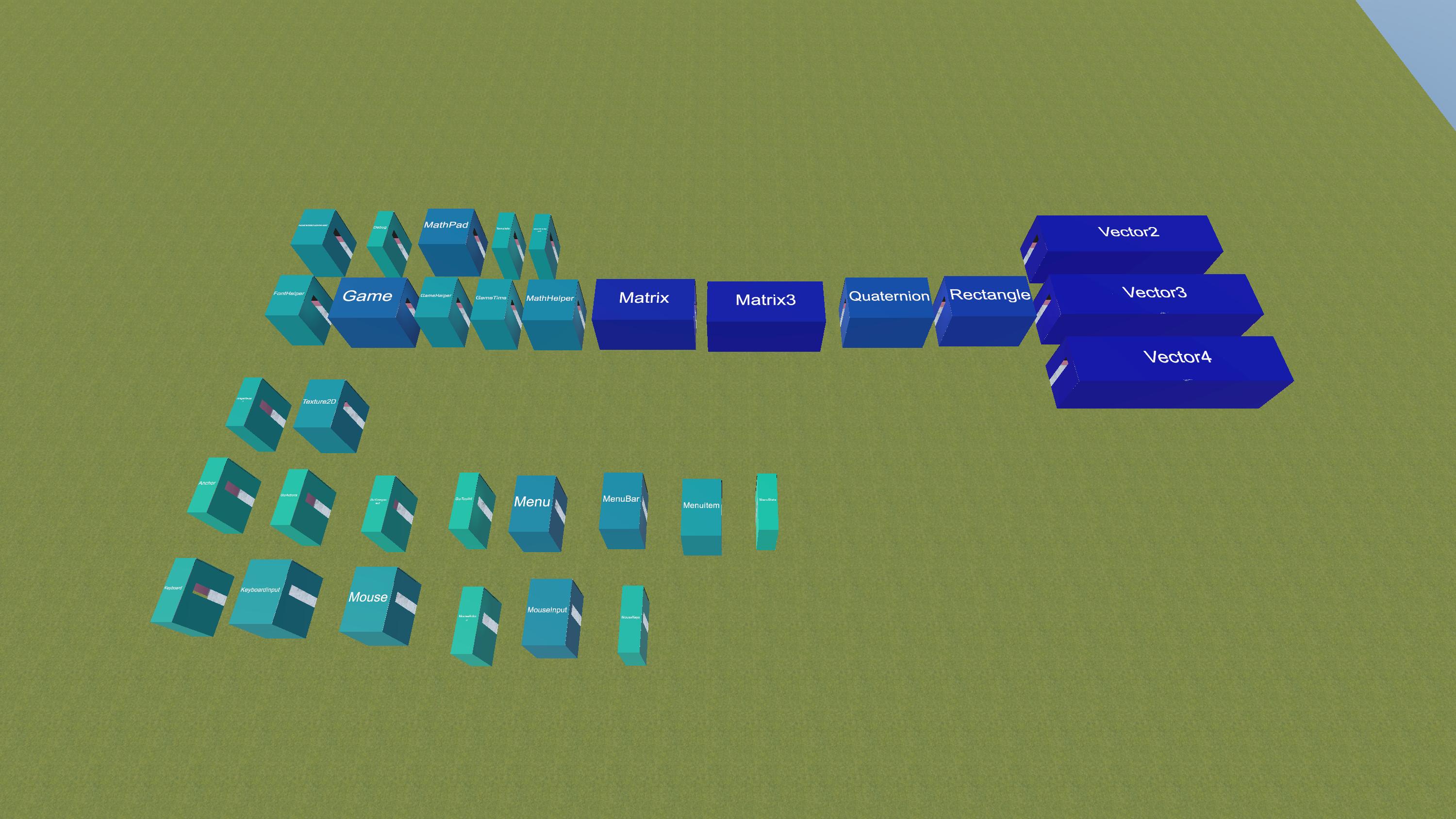}
  \caption{Participant 4} \label{fig:screenGame4}
\end{subfigure}
\begin{subfigure}{0.3\textwidth}
  \includegraphics[width=\columnwidth]{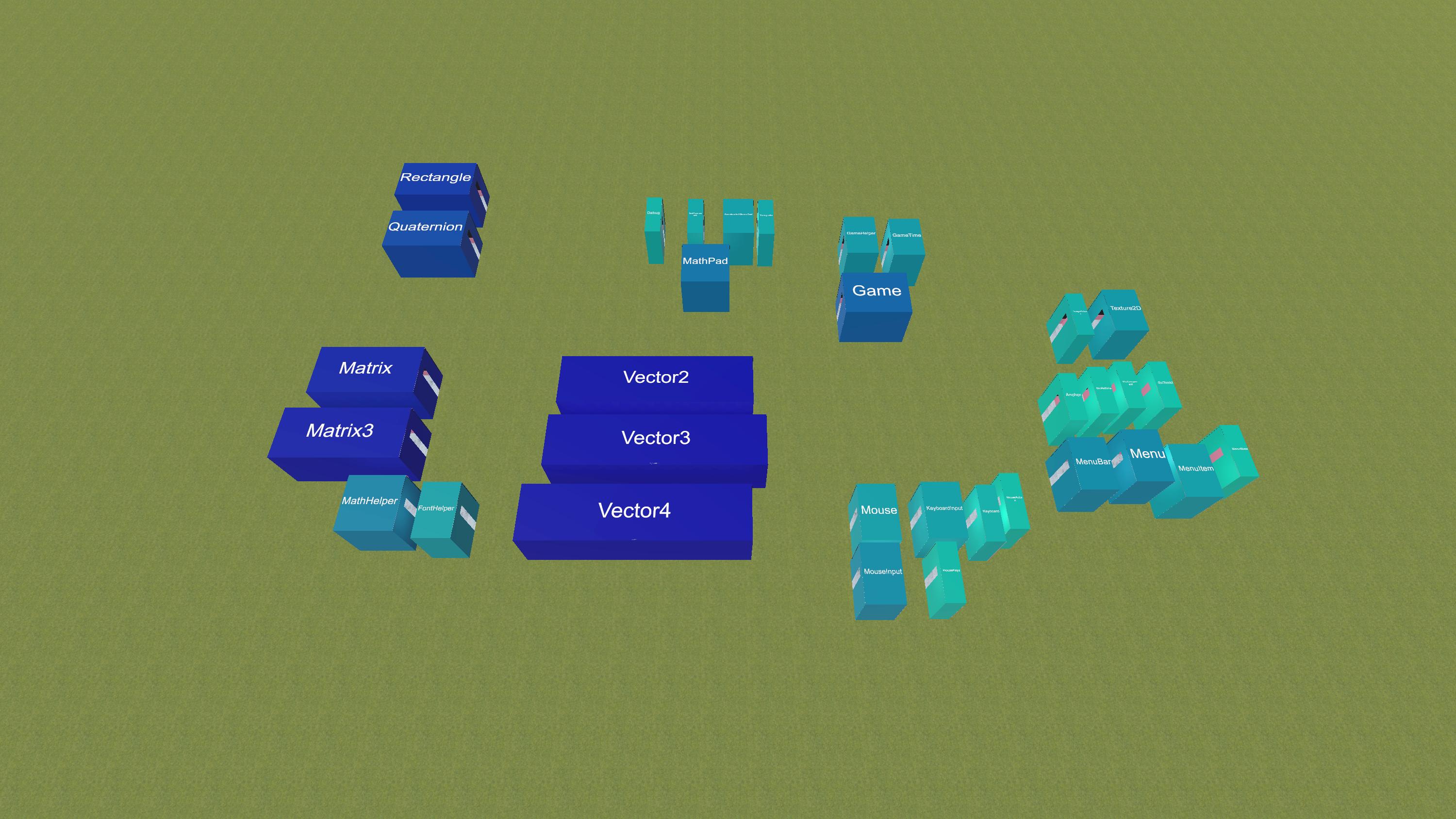}
  \caption{Participant 5} \label{fig:screenGame5}
\end{subfigure}
\begin{subfigure}{0.3\textwidth}
  \includegraphics[width=\columnwidth]{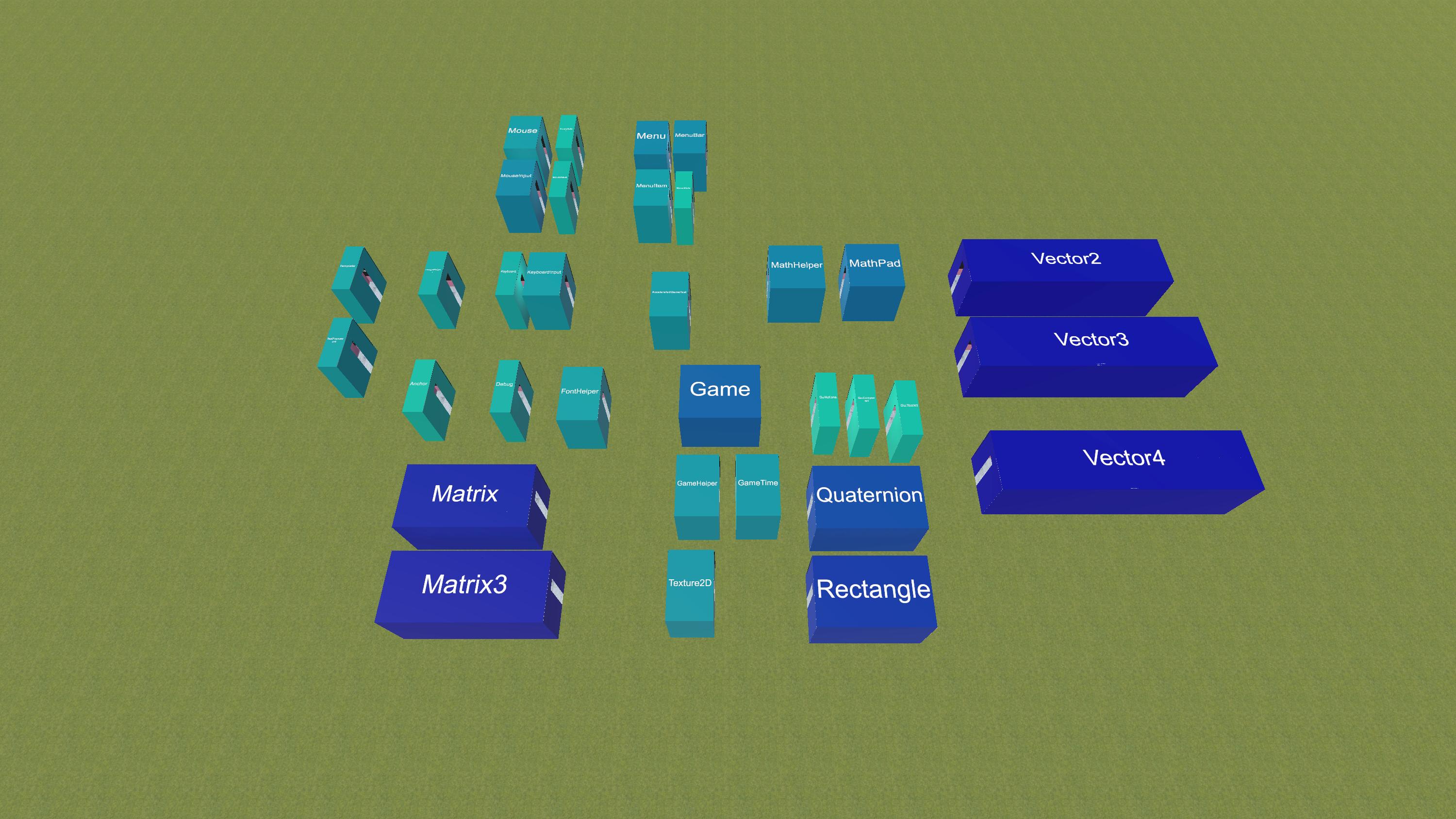}
  \caption{Participant 6} \label{fig:screenGame6}
\end{subfigure}
\begin{subfigure}{0.3\textwidth}
  \includegraphics[width=\columnwidth]{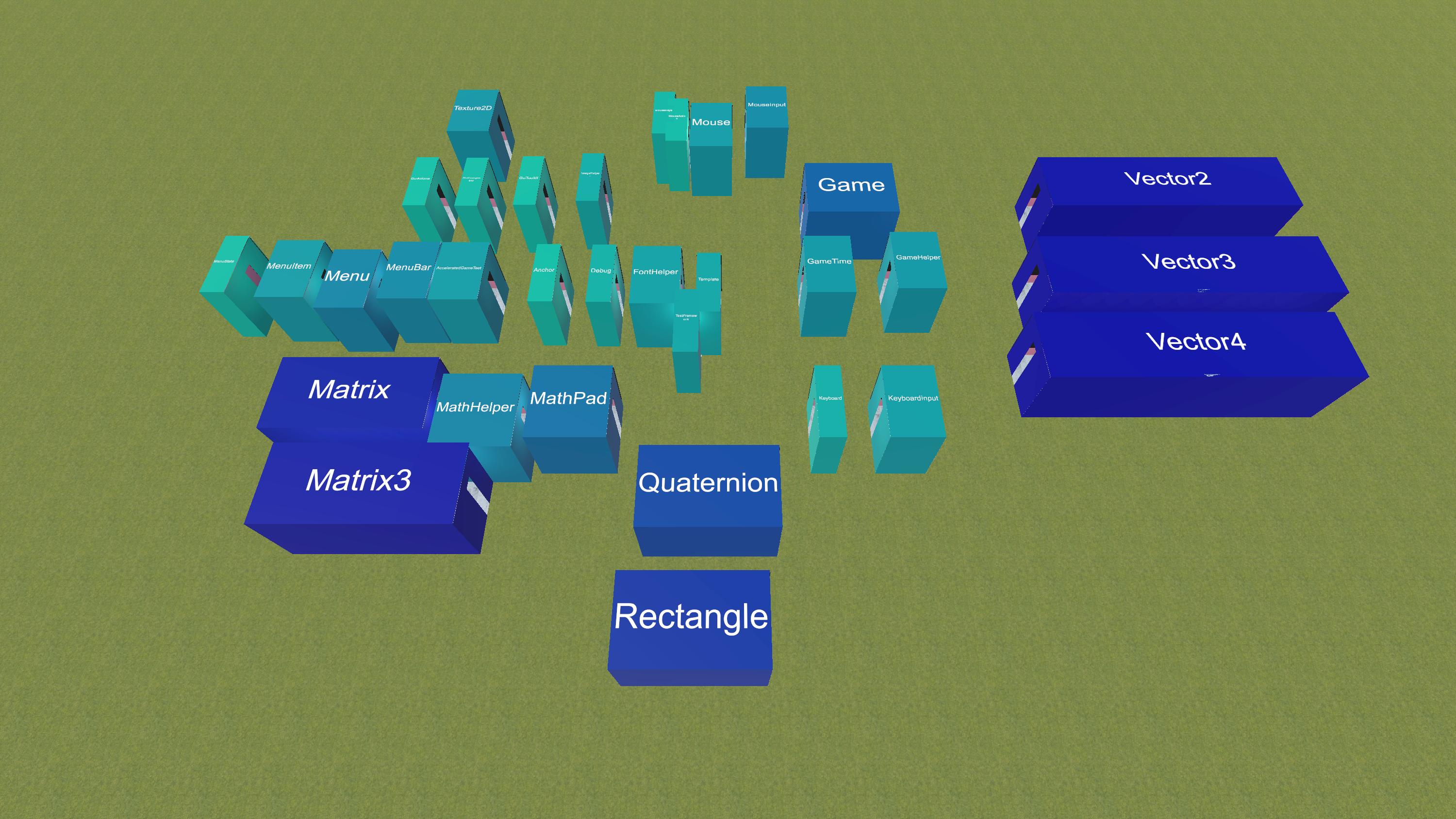}
  \caption{Participant 7} \label{fig:screenGame7}
\end{subfigure}
\begin{subfigure}{0.3\textwidth}
  \includegraphics[width=\columnwidth]{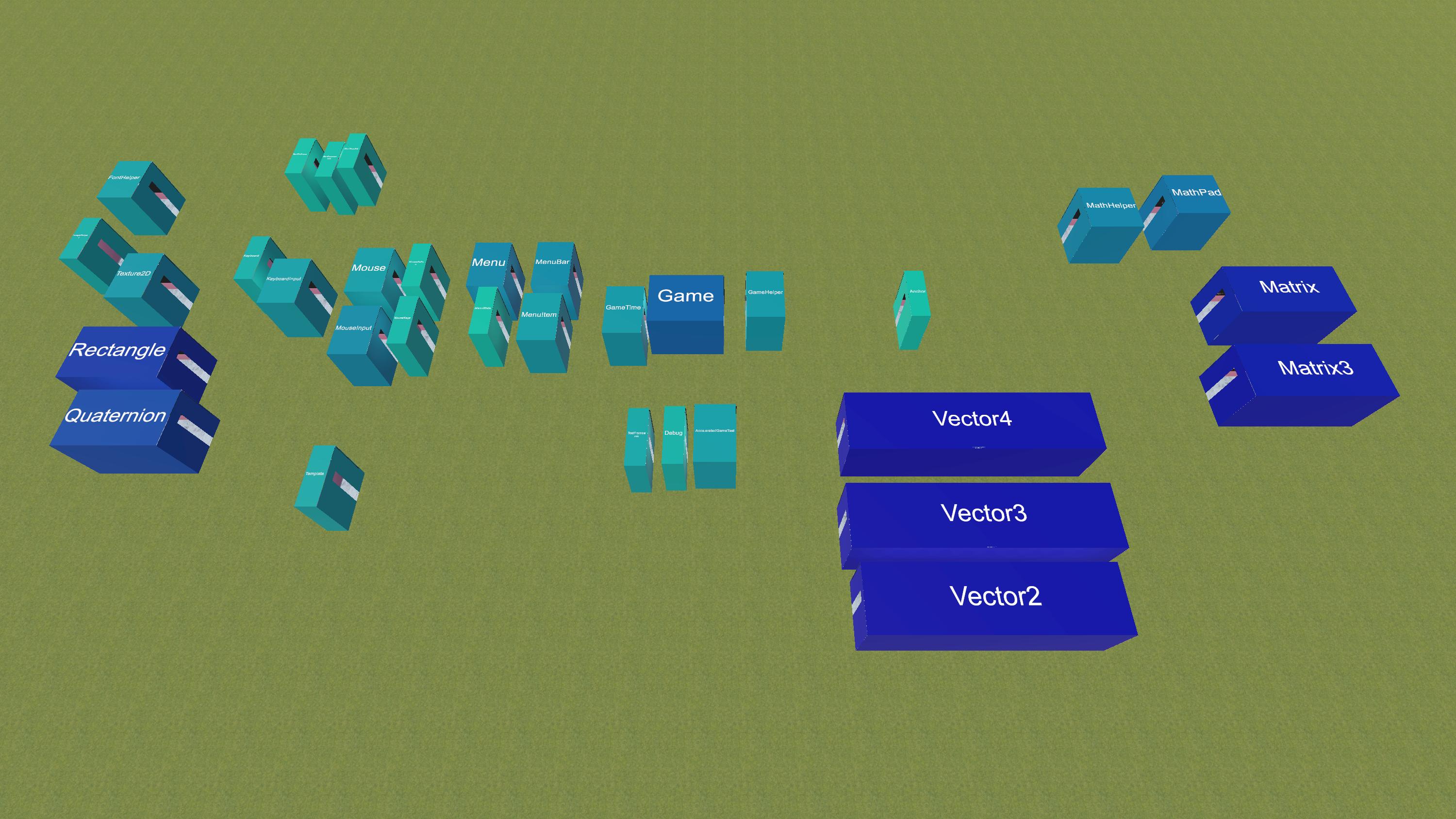}
  \caption{Participant 8} \label{fig:screenGame8}
\end{subfigure}
\begin{subfigure}{0.3\textwidth}
  \includegraphics[width=\columnwidth]{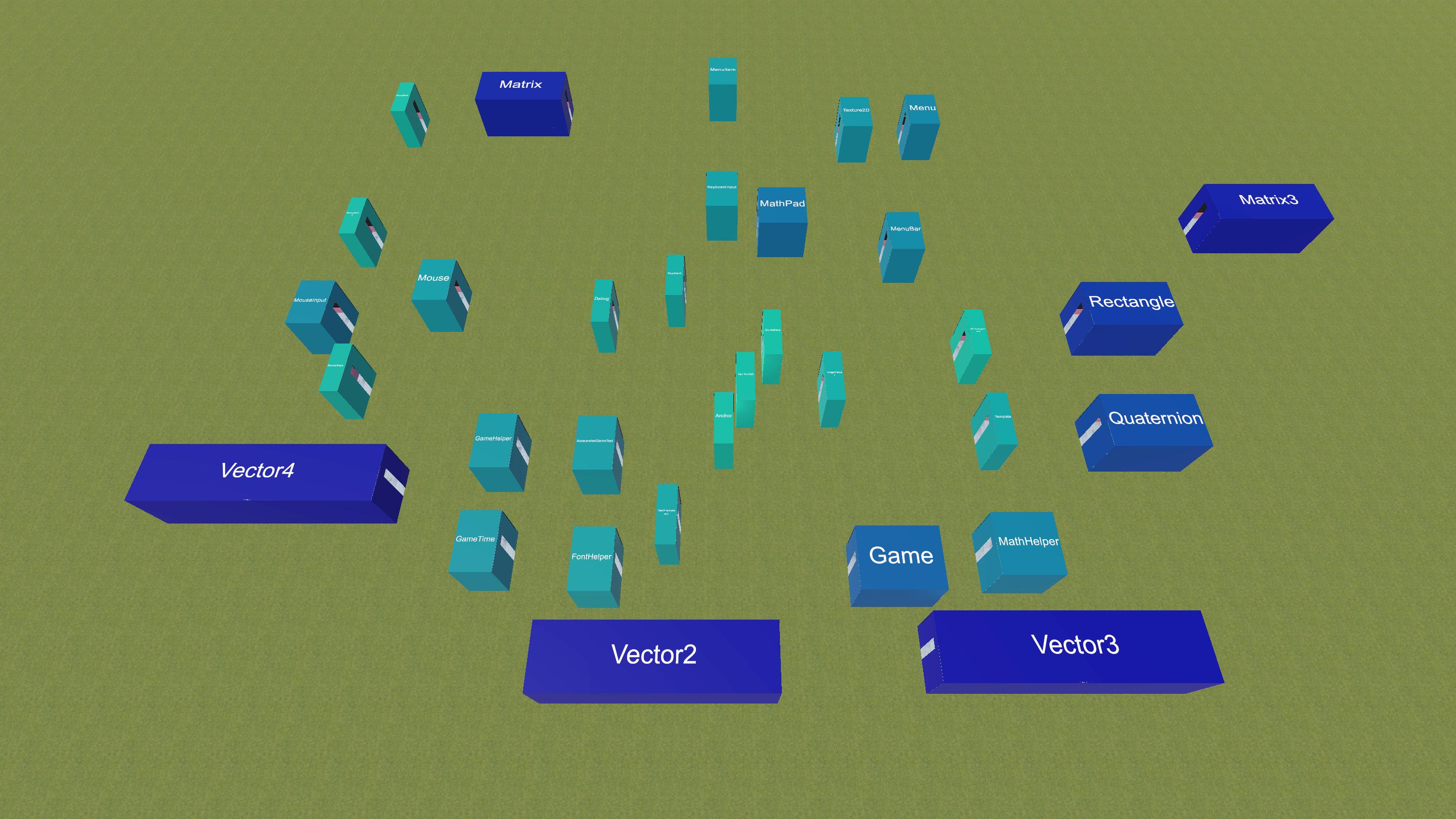}
  \caption{Participant 9} \label{fig:screenGame9}
\end{subfigure}
\caption{Screenshots of participants project organization.}
 \label{fig:screenGame}
\end{figure*}

Another possible explanation for observing faster task completion times with VS is that, CP provides more interaction capabilities compared to VS and 3D interactions are inherently slower compared to 2D interactions because of the added degree of freedom. When a user is inside a code room in CP, they have the freedom of walking around, looking at the wallpapers or clicking on the wallpapers to inspect the code more closely. All these interactions take time. We also gained some insight about this issue from a different perspective. Often times, after we assigned a task to a participant, we noticed that some participants started to ``play" with the interface or wander about aimlessly for a few seconds: due to the game-like environment of CP, they occasionally walked around the room and inspected the visual aspects of the environment, or would make a verbal note about something in the environment that was irrelevant to the assigned task (such as how realistic the reflection effect was on the floor tiles, or how the grass texture looked unrealistic).

Focusing on the quantitative results based on the experience level of the participants, reveals other findings. As one would expect, and regardless of the tool, beginners generally took more time to finish their tasks compared to experts. The other result of interest is the observed interaction effect between the tool and the experience of the participants. Figure \ref{fig:ChartInteractionTaskTimeTool_LVLEdited} presents the average time spent on determining the relationship between two classes (T3) based on each tool and the participants' experience level. The increase in the average time spent by the beginners when they switched from VS to CP was much more than this increase for experts. Without a more detailed study, it is difficult to draw any concrete conclusions. Informally, we think a possible explanation could be that it takes more time for the beginners to realize they do not know how to tackle T3. As a result, they take their time and explore the code further with CP, hoping to find a clue that aids them in completing the task.

Considering the codebase aspect of our study and focusing on the quantitative results, we see that the choice of codebase only affected the time spent for completing T3 (determining the relationship between two classes) and T5 (pinpointing a reasonable location in the code for adding the necessary logic to support some feature). This observation can be explained by noting the difference in the sizes of the two codebases. As detailed in Table \ref{tab:codebases}, MG is larger and more cluttered than LM. For T5, where the participants needed to study the code more closely, it generally took them longer to browse through MG compared to LM. Other than task completion time, the choice of codebase did not significantly affect the qualitative results. This bolsters our initial assumption that the two codebases were very similar and the choice of the codebase, would not significantly affect our results.

%----------------------------------------------------------------------------------------
%	Second Evaluation
%----------------------------------------------------------------------------------------

\section{Evaluation: Understanding Project Organization}
With the goal of determining the suitability of CP in the task of organizing an existing project, we designed and performed a second user study. In this user study, the users are tasked with organizing an existing project in Code Park in any way they saw fit.

%\subsection{Participants and Experience Design}
\vspace{2mm}
\noindent
\textbf{Participants and Experience Design.~}
We recruited 9 participants (8 males and 1 female ranging in age from 18 to 29 with a mean age of 22.8). Our requirements for participants were similar to the previous study, i.e. familiarity with the C\# language. Each participant was compensated with \$10 for their efforts at the end of the study session. Once again, each participant was given a pre-questionnaire containing some demographic questions as well as some questions asking about their experience in developing C\# applications.

Each participant was tasked with placing all 33 classes of an existing project in CP's environment in a manner that they saw reasonable. As a result, they were free to organize the classes in any way they preferred. We separated our participants into two groups of 5 and 4.

The classes in the project that was given to the first group were already organized into directories based on their relation (e.g. classes that handled user input were all inside of a directory called ``input"). Conversely, the classes in the project that was given to the second group were not organized in any particular manner (i.e. all classes were inside the same directory). The goal of such separation was to observe whether grouping the classes based on their inherent relationships would affect users' decisions.

\subsection{Results}

As evident in Figure \ref{fig:screenGame1} to Figure \ref{fig:screenGame5}, the organization performed by the participants mostly followed the directory-based organization of the project that was given to them. Some participants chose to place the contents of each directory in a separate line while others chose to spatially group them into a group of adjacent blocks. We asked the participants about their reasoning for such arrangements and obtained the following responses:

\begin{quote}
\textbf{Participant 1}: \textit{``Folders were arranged spatially in groups. Classes that appeared related by name were sub-grouped."}

\textbf{Participant 2}: \textit{``[I kept] directories grouped together."}

\textbf{Participant 3}: \textit{``I just arranged classes of a particular folder in each row."}

\textbf{Participant 4}: \textit{``The classes were arranged alphabetically for each folder and I arranged the classes in the same folder in the same line."}

\textbf{Participant 5}: \textit{``I tried to group the related class together based on the usefulness and field."}

\end{quote}

\vspace{2mm}
Figure \ref{fig:screenGame6} to \ref{fig:screenGame9} depicts the results obtained from the second group of participants. When asked about their reasoning for their decisions, the following responses were obtained:
\vspace{2mm}

\begin{quote}
\textbf{Participant 6}: \textit{``When arranging the classes my first concern was to group similar classes together. After considering which groups existed, I tried to come up with a hierarchy based on the classes size. So I put bigger classes on the side and all the smaller ones in the middle."}

\textbf{Participant 7}: \textit{``I tried to place the rooms in the chunk of similar classes. My priority was to place them in such a way that they are easy to find again."}

\textbf{Participant 8}: \textit{``I grouped the rooms based on their classes' name."}

\textbf{Participant 9}: \textit{``Big models together. Smaller ones in the middle so I can find them easier."}

\end{quote}

%----------------------------------------------------------------------------------------
%	DISCUSSION
%----------------------------------------------------------------------------------------

\subsection{Discussion}
In this user study our goal was to determine how users organize a project in CP environment in a way that the final result helped them remember the location of each class. The results show a possible relation between the user's cognitive understanding of the codebase and their decisions in organizing building block of the project when working with CP. As shown in Figure \ref{fig:screenGame}, users mostly chose to organize the constituent parts of the project based on their relationship with respect to each other. 

In cases where the project files were already organized into directories, users mostly followed that same organization when working with CP. However, if the project lacked an inherent organization, users' decisions were guided either by the size of each class or the semantic relationship of those classes. Users mostly elected to organize similar parts of the project in the close proximity of each other. This is inline with the results observed in \cite{abbes} where the increase in spatial dispersion of objects resulted in more difficulty in processing and attentional allocation. 

\section{Limitations and Future Work}
There are a few notable limitations associated with the design of our first study and CP in general. First, we realize that our comparison with VS could potentially bias the results. This is because most C\# developers have experience with VS and such prior familiarity could affect their responses. Second, comparing VS which is fast and responsive to an interface that has animations and is slower may not result in a completely fair assessment.

Another limitation is that we compared VS against CP on a strict code understanding basis. Compared to VS, CP misses code editing or debugging functionalities. Also, CP currently only supports C\# codebases. However, it can be easily extended to any object-oriented programming language. Given these limitations, our results and the participants' feedback indicate a trend in preference for CP and tools that drastically change the way a programmer interacts with the code. Our results show that CP is at least as good as a professional tool such as VS in learning a codebase. 

We plan to address these limitations by incorporate more functionality into CP such as a code editor and a debugger. These were the most requested features by our user study participants. We also would like to perform a more in-depth study to observe how CP would affect the learning of a group of novice programmers in a semester long course similar to the work of Saito \textit{et al.} \cite{education}. Zorn \textit{et al.} \cite{minecraft} examined the game Minecraft as a means of increasing interest in programming. It would be interesting to perform a similar study on CP and evaluate its effectiveness on programming.

Beyond system hardware limits, we believe CP can scale, through extended metaphors. Since a world of buildings will likely become incomprehensible, we expect a project to grow from buildings to districts, to cities, to regions, and so on. Understanding the boundary and limit of each is future work. We also plan to evaluate CP in virtual reality (VR) and augmented reality (AR) environments. It would be interesting to design an AR system that employs CP to aid in code understanding while allowing programmers to naturally use the mouse and the keyboard.

%----------------------------------------------------------------------------------------
%	CONCLUSION AND FUTURE WORK
%----------------------------------------------------------------------------------------
\section{Conclusion}
We presented Code Park, a 3D code visualization tool designed to improve learning a codebase. Code Park lays out an existing codebase in a 3D environment and allows users to explore and study the code in two modalities, a bird's eye view mode and a first-person view mode. We performed a user study to evaluate the usability and the effectiveness of Code Park in comparison to Microsoft Visual Studio in performing a series of task related to the user's understanding of the codebase. We then performed a follow up user study to evaluate how users would organize an existing project in the 3D environment in a manner that would help them remember the codebase later. The analysis of our results demonstrated the benefits of Code Park as a viable tool for understanding an existing codebase. Our participants found Code Park to be easy to learn, instrumental in understanding as well as remembering the structure of a codebase, and enjoyable to use. Our analysis of the first study revealed that the participants that did not have a strong programming background found Code Park to be easier to work with compared to Microsoft Visual Studio. The result of second study showed that the users tend to organize project in a semantically meaningful form.

%----------------------------------------------------------------------------------------
%	Acknowledgements
%----------------------------------------------------------------------------------------
\section{Acknowledgements}
This work is supported in part by NSF Award IIS-1638060, Lockheed Martin, Office of Naval Research Award ONRBAA15001, Army RDECOM Award W911QX13C0052, and Coda Enterprises, LLC. We also thank the ISUE lab members at UCF for their support as well as the anonymous reviewers for their helpful feedback.

\bibliographystyle{IEEEtran}

\bibliography{sample}
\end{document}